\documentclass[aps,prd,reprint,superscriptaddress,floatfix]{revtex4-2}

\usepackage{graphicx}
\usepackage{amsmath,amssymb,amsthm,amsfonts}
\usepackage{bbold}
\usepackage{float}
\usepackage{slashed}
\usepackage{dcolumn}
\usepackage{bm}
\usepackage{color}
\usepackage{multirow}
\usepackage{fontawesome}
\usepackage{soul}

\usepackage{hyphenat}
\usepackage[pdfencoding=auto]{hyperref}
\usepackage[final]{microtype}

\def\bea{\begin{eqnarray}}
\def\eea{\end{eqnarray}}
\def\beq#1\eeq{\begin{align}#1\end{align}}
\def\beqnn#1\eeq{\begin{align*}#1\end{align*}}
\def\ba{\begin{array}}
\def\ea{\end{array}}
\def\bc{\begin{center}}
\def\ec{\end{center}}

\def\gev{\rm GeV}

\usepackage{breqn} 

\usepackage{cases}

\newcommand{\be}{\begin{equation}\begin{aligned}}
\newcommand{\ee}{\end{aligned}\end{equation}}

\newcommand{\beqa}{\begin{eqnarray}}
\newcommand{\eeqa}{\end{eqnarray}}

\renewcommand{\eqref}[1]{Eq.~(\ref{#1})}

\newcommand{\eg}{{\em e.g.}}

\DeclareUnicodeCharacter{2212}{\textendash}

\makeatletter
\def\l@subsubsection#1#2{}
\makeatother

\newcommand{\tr}{\operatorname{Tr}}

\usepackage{makecell}

\usepackage{cleveref}
\crefname{section}{Sec.}{Secs.}
\crefname{figure}{Fig.}{Figs.}
\crefname{equation}{Eq.}{Eqs.}
\crefname{appendix}{Appendix}{Appendices}

\begin{document}

\preprint{CPTNP-2026-009}

\title{Impact of flavor changing processes on prospects \\for majoron discovery at intensity-frontier searches}

\author{Krzysztof Jod\l{}owski}
\email{Corresponding author: 91058@njnu.edu.cn}

\affiliation{Department of Physics and Institute of Theoretical Physics Nanjing Normal University\char`,{} Nanjing\char`,{} 210023\char`,{} China}
\affiliation{Nanjing Key Laboratory of Particle Physics and Astrophysics\char`,{} Nanjing\char`,{} 210023\char`,{} China}
\affiliation{Particle Theory and Cosmology Group\char`,{} Center for Theoretical Physics of the Universe\char`,{}  Institute for Basic Science (IBS)\char`,{} Daejeon 34126\char`,{} Korea}

\author{Chih-Ting Lu}
\email{ctlu@njnu.edu.cn}
\affiliation{Department of Physics and Institute of Theoretical Physics Nanjing Normal University\char`,{} Nanjing\char`,{} 210023\char`,{} China}
\affiliation{Nanjing Key Laboratory of Particle Physics and Astrophysics\char`,{} Nanjing\char`,{} 210023\char`,{} China}

\begin{abstract}
  The singlet majoron $J$ is the pseudo-Nambu-Goldstone boson of a spontaneously broken global $U(1)_{B-L}$ symmetry that generates Majorana masses for right-handed neutrinos.
  Its tree-level neutrino coupling scales as $m_\nu/f_J$, where $f_J$ is the symmetry-breaking scale, while charged-fermion and gauge-boson couplings, including flavor-changing ones, arise at loop level. 
  Consequently, $J$ can be long-lived over wide regions of parameter space, motivating displaced-decay searches. 
  We study these signatures at proton beam dumps, neutrino facilities, the LHC far detectors, and $Z$ boson factories (including DUNE, NA62, FASER/FASER2, MATHUSLA, SHiP, CEPC, and FCC-ee). 
  We investigate representative positive-semi-definite anarchical, single-flavor, and CP-violating coupling textures, showing that $\tau^\pm\to\ell^\pm J$ decays, where $\ell=e,\mu$, dominate production in the mass range $m_J\simeq0.2\text{--}1.7\,\mathrm{GeV}$.
  We find that the combined constraints from BEBC, CHARM, and NuCal are stronger than those from ArgoNeuT, with CEPC and FCC-ee providing a complementary sensitivity to existing flavor-violation searches. 
  Finally, SHiP is expected to provide the strongest constraints, probing effective coupling scales as high as $f\sim 10^9\,\mathrm{GeV}$.
\end{abstract}


\maketitle

\section{Introduction}
\label{sec:introduction}
The observation of neutrino oscillations has established that at least two neutrinos are massive and that lepton flavors mix, providing unambiguous evidence for physics beyond the Standard Model (SM). Among the most economical and theoretically appealing frameworks to account for these masses is the type-I seesaw mechanism~\cite{Minkowski:1977,Gell-Mann:1979vob,Yanagida:1979as,Mohapatra:1979ia}, in which SM-singlet right-handed neutrinos acquire large Majorana masses, and light neutrino masses arise from the exchange of these heavy states. 

In many realizations, the seesaw is embedded in a theory with a global, non-anomalous $U(1)_{B-L}$ symmetry, spontaneously broken by the vacuum expectation value (VEV) of a scalar field.
The Nambu-Goldstone boson is the majoron~\cite{Chikashige:1980ui,Gelmini:1980re,Schechter:1981cv}, a pseudoscalar associated with neutrino mass generation. 
Therefore, unlike a generic axion-like particle (ALP) with arbitrary couplings, the majoron's interactions depend on the seesaw parameters.
However, while its interactions with charged leptons are correlated, they are not fixed by neutrino oscillation data.
Indeed, these couplings depend on $M_D M_D^\dagger$, while the light-neutrino mass matrix involves $M_D M_N^{-1}M_D^T$; see~\cref{eq:mass_matrices,eq:seesaw} below.
Consequently, these couplings depend on the specifics of an unknown UV completion, which motivates studying a set of complementary coupling textures and their experimental consequences.

The singlet-majoron type-I seesaw provides a reference realization of the coupling structure relevant to our phenomenological analysis -- we first recall this construction and introduce the effective interactions.
The SM is augmented by a complex scalar singlet $S$ carrying $B\!-\!L$ charge $2$, so that its VEV generates the right-handed neutrinos (RHN) Majorana masses,
\be
    \mathcal{L} \supset - y_{\alpha i}\,\overline{L}_\alpha \tilde{H} N_i - \frac{1}{2} \lambda_{ij}\, S\, \overline{N_i^c} N_j + \mathrm{h.c.} - V(H, S)
  \,,
\label{eq:L_seesaw}
\ee
where $L_\alpha$ ($\alpha = e, \mu, \tau$) denote the SM lepton doublets, $\tilde{H} = i\sigma_2 H^*$ is the conjugate Higgs doublet, $N_i$ ($i=1,\ldots,n_N$,  $n_N\geqslant 3$) are SM-singlet RHNs, $y_{\alpha i}$ are the neutrino Yukawa couplings, and $\lambda_{ij} = \lambda_{ji}$ parameterize the Majorana Yukawa interactions. 
The scalar potential $V(H,S)$ respects the global $U(1)_{B-L}$ symmetry under which leptons transform with charge $-1$: $L_\alpha \to e^{-i\theta} L_\alpha$, $N_i \to e^{-i\theta} N_i$, and $S \to e^{2i\theta} S$, while quarks transform with charge $1/3$.

Spontaneous symmetry breaking occurs when both scalars acquire VEVs,
\be
    \langle H \rangle = \frac{1}{\sqrt{2}}\begin{pmatrix} 0 \\ v \end{pmatrix}, \qquad
    \langle S \rangle = \frac{f_J}{\sqrt{2}}\,,
\label{eq:vevs}
\ee
with $v \simeq 246\,\mathrm{GeV}$ the electroweak (EW) scale and $f_J \gg v$ the $B\!-\!L$ breaking scale.
These generate the Dirac and Majorana mass matrices
\be
  (M_D)_{\alpha i} = \frac{y_{\alpha i}\, v}{\sqrt{2}}\,, \qquad
  (M_N)_{ij} = \frac{\lambda_{ij}\, f_J}{\sqrt{2}}
\,.
\label{eq:mass_matrices}
\ee
Integrating out the heavy right-handed neutrinos yields the light neutrino mass matrix via the seesaw formula,
\be
    m_\nu = - M_D\, M_N^{-1}\, M_D^T\,,
\label{eq:seesaw}
\ee
which relates the light-neutrino masses to the Yukawa matrices and the heavy-neutrino mass scale.

The complex scalar $S$ can be parameterized as
\be
  S(x) = \frac{1}{\sqrt{2}}\bigl(f_J + \rho(x)\bigr)\exp\left[i\,\frac{J(x)}{f_J}\right]
  \,,
\label{eq:S_param}
\ee
where $\rho(x)$ is the massive radial mode and $J(x)$ is the massless Nambu--Goldstone boson associated with the spontaneous breaking of $U(1)_{B-L}$---the majoron. 
The radial mode $\rho$ acquires a mass $m_\rho \simeq \sqrt{2\lambda_S}\,f_J$, where $\lambda_S$ is the quartic self-coupling of $S$ in $V(H,S)$, and it decouples from the low-energy phenomenology, whereas the majoron remains in the spectrum as a light degree of freedom with derivative couplings suppressed by $1/f_J$.

In the exact $U(1)_{B-L}$ limit, the majoron is strictly massless and couples derivatively to the lepton-number current. However, exact continuous global symmetries are not expected to survive in a UV-complete theory of quantum gravity~\cite{Giddings:1989bq,Harlow:2018jwu,Harlow:2018tng}. 
Planck-suppressed operators and possibly nonperturbative gravitational effects, such as Euclidean wormholes, explicitly break $U(1)_{B-L}$~\cite{Giddings:1987cg,Coleman:1988cy,Abbott:1989jw}, lifting the flat direction of the scalar potential. 
Additional sources of explicit breaking include higher-dimensional operators suppressed by a UV cutoff. 
These effects induce a non-zero mass, $m_J \neq 0$, promoting the majoron to a pseudo-Nambu-Goldstone boson. 
The resulting low-energy effective Lagrangian for the majoron field takes the form
\be
    \mathcal{L}_J = \frac{1}{2}\partial_\mu J\, \partial^\mu J - \frac{1}{2} m_J^2\, J^2 + \mathcal{L}_{\mathrm{int}} 
\,,
\label{eq:L_kinetic}
\ee
where $m_J$ parameterizes the strength of explicit breaking, and which we treat as a free parameter due to the lack of detailed knowledge of the relevant UV physics. 
We also assume that in the parameter space of interest explicit $U(1)_{B-L}$ breaking manifests itself as the mass term $m_J$, while additional shift-symmetry-violating operators affecting $J$ couplings are negligible.

A majoron with mass in the MeV--GeV range has received considerable recent attention, with studies demonstrating its potential impact on Big Bang Nucleosynthesis through modifications to primordial light-element abundances. 
However, the updated BBN bound on the majoron is weak, $1/f_J\simeq g/m_\nu \sim 0.2/\mathrm{GeV}$~\cite{Chang:2024mvg}. 
Astrophysical constraints, particularly from supernovae, can be stringent~\cite{Choi:1987sd,Fiorillo:2023,Akita:2023iwq,Li:2025beu,Huang:2025rmy,Huang:2025xvo}, though recent analyses have shown that in the strong-coupling (trapping) regime the impact on the supernova neutrino signal is considerably more limited than previously estimated~\cite{Fiorillo:2023ytr,Fiorillo:2023cas}, and that observables beyond the standard luminosity-based cooling argument can provide the dominant constraints, largely superseding earlier cooling bounds~\cite{Fiorillo:2025sln}. 
A future Galactic supernova would significantly extend this reach~\cite{Telalovic:2024cot}. 
These bounds are typically restricted to $m_J \lesssim \mathcal{O}(0.1\,\mathrm{GeV})$, while above this threshold, constraints weaken appreciably. 

The \textit{intermediate} window $0.1\,\mathrm{GeV} \lesssim m_J \lesssim 2\,\mathrm{GeV}$ therefore remains comparatively underexplored: supernova cooling bounds lose sensitivity once the majoron mass exceeds the core temperature, while in the higher-mass part of the parameter space, $B$-meson factories and dedicated lepton-flavor-violating (LFV) searches provide leading coverage~\cite{Heeck:2019guh,Cheng:2020rla,Escribano:2021uhf,Herrero-Brocal:2023czw}.
While a recent analysis~\cite{Bertuzzo:2022fcm} of ArgoNeuT~\cite{Anderson:2012vc} data has obtained bounds on sub-GeV majorons, the coverage remains incomplete, and a more systematic study of intensity-frontier (IF) experiments is needed to improve it.

In this work, we study the majoron in IF experiments, focusing on the intermediate-mass regime $m_J \lesssim m_\tau \simeq 1.8\,$GeV.
We demonstrate that the LFV decays, $\tau^\pm\to\ell^\pm J$, where $\ell=e,\mu$, provide the leading majoron production channels in this window, and using Monte Carlo (MC) simulation, we assess the resulting sensitivity at beam dump experiments, neutrino facilities, forward detectors at the LHC, and $Z$ boson factories.
Our emphasis is on a systematic assessment of their experimental impact across positive-semidefinite effective coupling textures, including anarchical benchmarks satisfying positivity constraints, single-flavor interactions, and complex benchmarks.

\section{Coupling structure}
\label{sec:Coupling_structure}
In the following, we discuss the full form of the interaction Lagrangian. 
As the Nambu-Goldstone boson of a global $U(1)_{B-L}$, $J$ couples derivatively to the $B\!-\!L$ current. 
A recent work~\cite{Herrero-Brocal:2026nmc} argues that if a single VEV breaks several global $U(1)$ combinations, some gauge-anomalous (e.g. $L$) and one gauge-anomaly free ($B\!-\!L$), the resulting Nambu-Goldstone boson necessarily aligns with the anomaly-free $B\!-\!L$, further motivating our choice.
Whenever we use results derived in lepton-number majoron notation in the literature~\cite{Quevillon:2019zrd,Heeck:2019guh}, we match onto the same low-energy effective couplings relevant for on-shell $J$ processes considered here.

Since the charged-lepton Dirac masses preserve $U(1)_{B-L}$, one can choose a basis in which the majoron field is removed from the charged-fermion mass/Yukawa terms by symmetry-preserving field redefinitions. Equivalently, any apparent tree-level $J$-coupling to charged SM fermions is a derivative coupling to a conserved current and does not lead to physical on-shell amplitudes. Physical couplings of $J$ to charged fermions arise only radiatively once EW symmetry is broken and the seesaw sector is integrated out.
Upon using the equations of motion, the only irreducible tree-level coupling to SM fields is $J\nu\nu$, which is proportional to the Majorana neutrino masses, and is negligible for our purposes. 
Similarly, couplings involving $J$ and heavy neutrinos do not affect majoron production or decay modes at the IF.

For ease of comparison, we adopt the conventions of Ref.~\cite{Bertuzzo:2022fcm}, while indicating the explicit matching to Ref.~\cite{Heeck:2019guh}. The tree-level coupling between the majoron and active neutrinos reads
\be
  \mathcal{L}^{\mathrm{tree}}_{\mathrm{int}} = -\frac{\partial_\mu J}{4f_J} \, \sum_{i=1}^3 \bar{\nu}_i \,\gamma^\mu \gamma_5 \,\nu_i = \frac{iJ}{2f_J}\sum_{i=1}^{3} m_i \,\bar{\nu}_i\, \gamma_5\, \nu_i 
\,,
\label{eq:L_tree}
\ee
where $f_J=\sqrt{2}\langle S\rangle$ is the physical singlet VEV and the second equality follows from the equations of motion. 
The effective scale $f$ used for the charged-fermion couplings below is defined with the benchmark textures, while the scale $f_J$ is sufficiently large that
\be
  \Gamma(J\to\nu\nu)=\frac{m_J}{16\pi f_J^2}\sum_i m_i^2
  \ll \Gamma_{\mathrm{vis}}\,,
\ee
where $\Gamma_{\mathrm{vis}}$ is the total visible decay width.

Radiative corrections at one~\cite{Pilaftsis:1993af,Broncano:2002rw,Garcia-Cely:2017oco} and two loops~\cite{Heeck:2019guh} induce interactions with charged fermions $\psi$ and gauge bosons as follows:
\be
  \mathcal{L}^{\mathrm{loop}}_{\mathrm{int}} &= \frac{1}{2f}\, \partial_\mu J \, \bar{\psi} \gamma^\mu \bigl( C_V + C_A \gamma_5 \bigr) \psi + \frac{g_{J\gamma\gamma}}{4} J\, F_{\mu\nu} \tilde{F}^{\mu\nu} \\
  &\quad+ \frac{g_{J\gamma Z}}{2} J\, F_{\mu\nu} \tilde{Z}^{\mu\nu}  + \frac{g_{Jgg}}{4} J\, G_{\mu\nu}^a \tilde{G}^{a\mu\nu} \\
  &\quad+ \frac{g_{JZZ}}{4} J\, Z_{\mu\nu} \tilde{Z}^{\mu\nu} + \frac{g_{JWW}}{4} J\, W_{\mu\nu}^+ \tilde{W}^{-\mu\nu} 
  \,.
\label{eq:L_1}
\ee
Here, $\tilde{V}^{\mu\nu} = \epsilon^{\mu\nu\rho\sigma}V_{\rho\sigma}/2$ denotes the field-strength tensor dual to $V^{\mu\nu}$.
As discussed in~\cite{Heeck:2019guh}, the photon and gluon operators are proportional to $\partial^2 J\, V\tilde{V}$ and vanish for massless majoron. 
Application of the equations of motion yields~\eqref{eq:L_1}, so for processes in which $J$ appears exclusively as an on-shell external state (as is the case throughout this work) the two forms are equivalent.
We neglect the impact of the flavor-violating (FV) quark currents, since they arise at two-loop order, and its precise phenomenological consequences depend on specifics of the heavy-neutrino parameters.
Additional couplings of $J$ to $\gamma H$ and $ZH$ (with $H$ the SM Higgs) have negligible impact for detection prospects of GeV-scale majoron.

A key role in majoron phenomenology is played by the dimensionless matrix $K$, which has the following form:
\be
  K = \frac{M_D M_D^\dagger}{v\,f_J} = \begin{pmatrix} K_{ee} & K_{e\mu} & K_{e\tau} \\ K^{*}_{e\mu} & K_{\mu\mu} & K_{\mu\tau} \\ K^{*}_{e\tau} & K^{*}_{\mu\tau} & K_{\tau\tau} \end{pmatrix}
  \,,
\label{eq:K_mat}
\ee
where $M_D$ is the Dirac mass matrix mixing active and sterile neutrinos, $v$ is the Higgs VEV.
The diagonal entries are real, while the off-diagonal ones can be complex.
The matrix $K$ has several important properties: it enters the expressions for the majoron couplings generated at loop level~\cite{Chikashige:1980ui,Garcia-Cely:2017oco}, being constructed from $M_D M_D^\dagger$, it is hermitian and positive semi-definite by construction, even when one light neutrino is massless~\cite{Davidson:2006tg}.
However, while neutrino oscillation data probe the combination $M_D M_N^{-1} M_D^T$ (cf.\ \eqref{eq:seesaw}), majoron couplings depend on $M_D M_D^\dagger$, which is not fixed by oscillation data.

\begin{figure}[tb]
      \centering
          \includegraphics[scale=0.38]{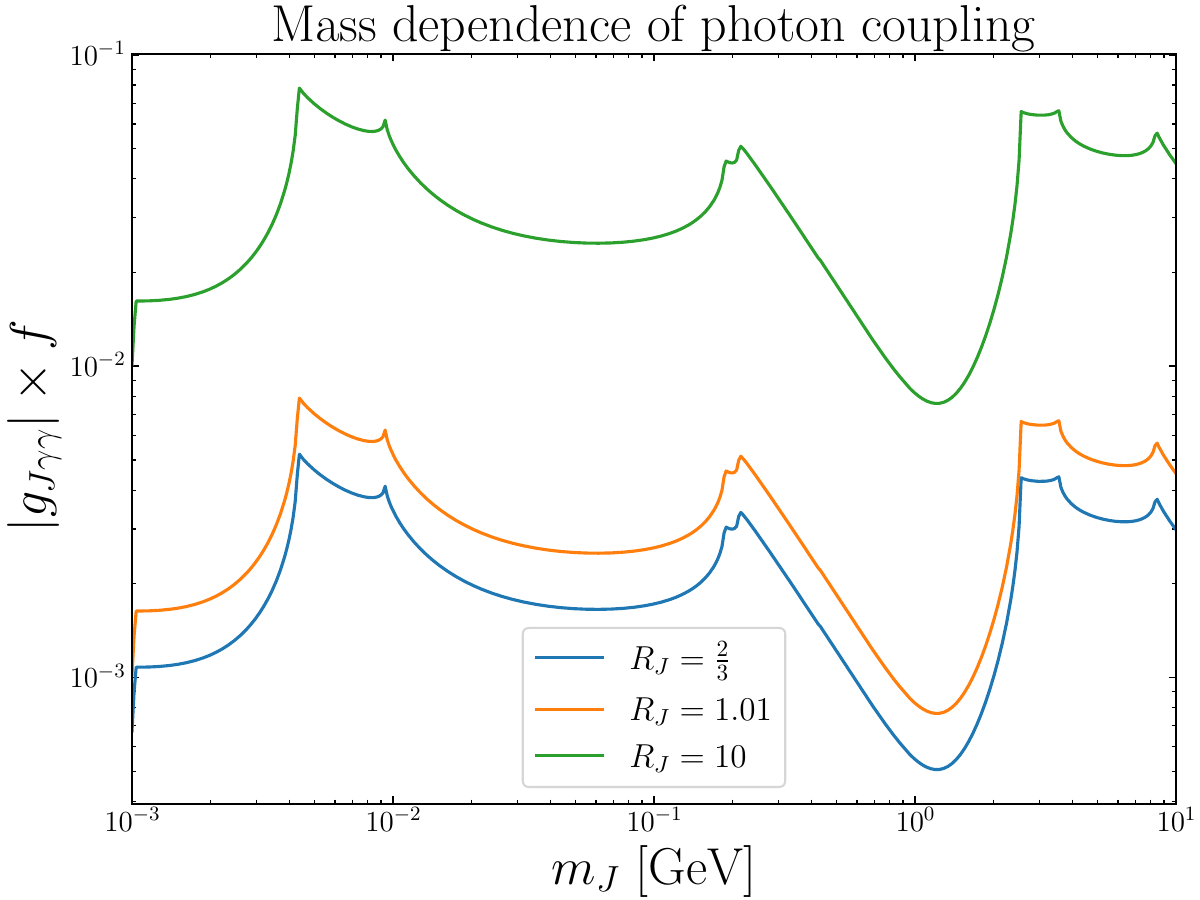}
          
          \hspace{0.1cm}

          \includegraphics[scale=0.38]{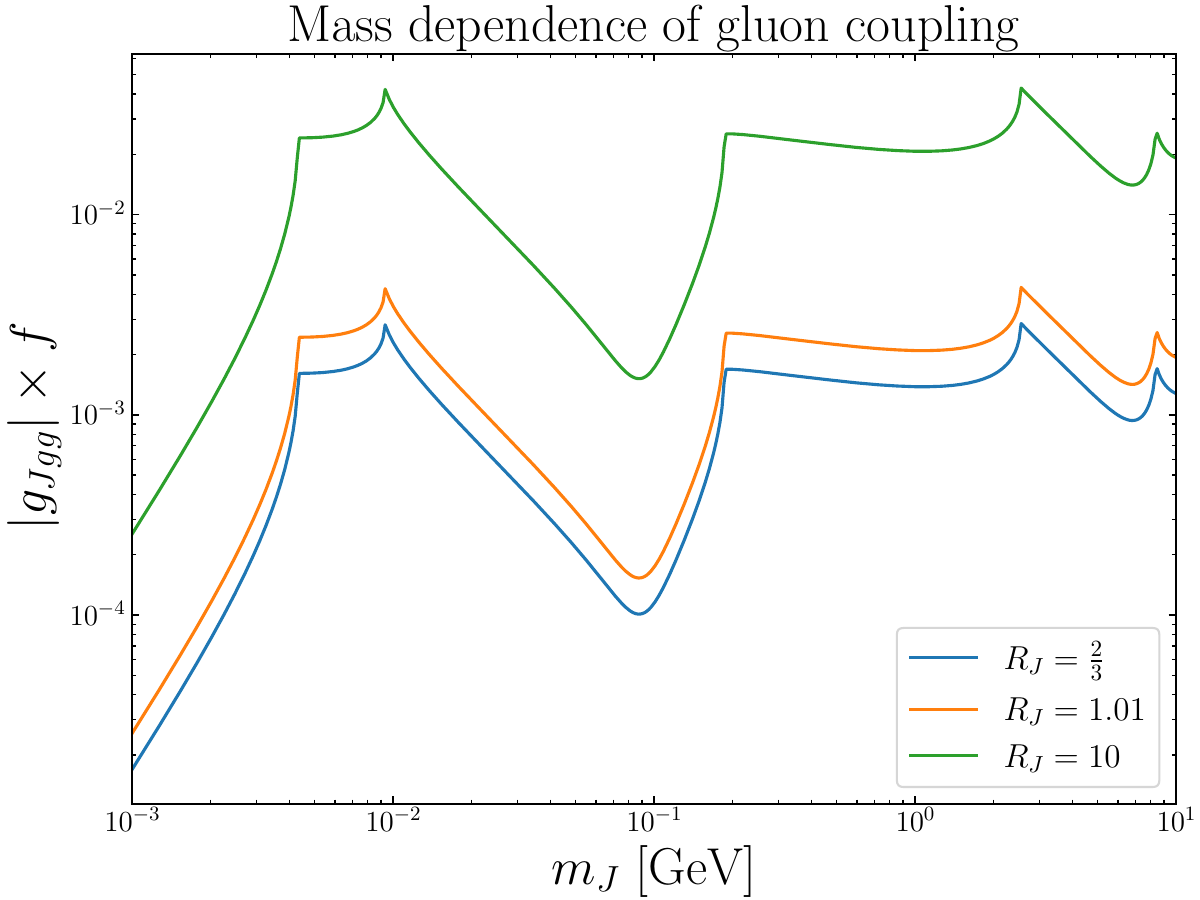}
          
          \hspace{0.1cm}
          
          \includegraphics[scale=0.38]{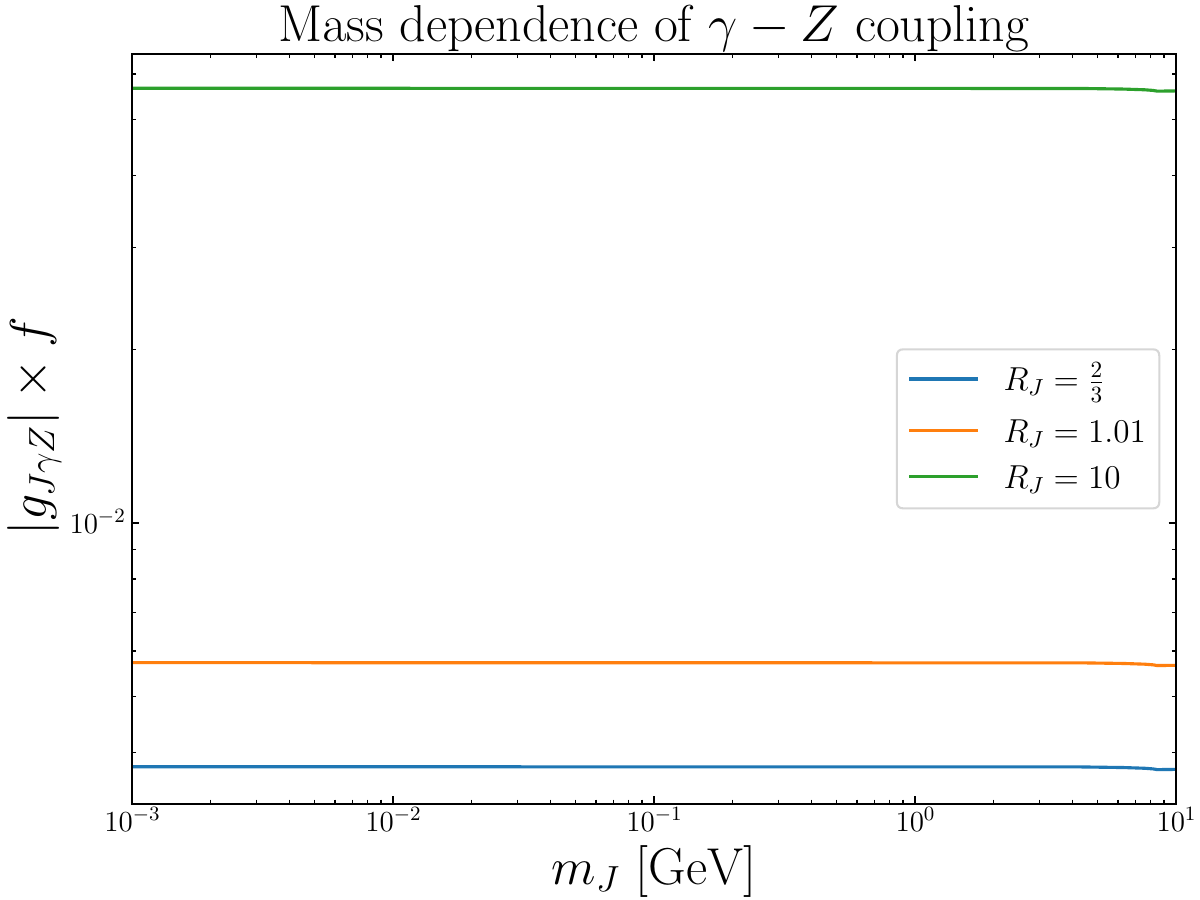}
          \caption{
                  Functional dependence of the majoron couplings to gauge bosons, which affect the decay widths and production modes of the majoron. 
                  We show three examples (indicated  by blue, orange, and green lines) of the ``anarchical'' benchmark defined by \eqref{eq:K_anarch}.
                  A similar dependence holds for the benchmarks given by \eqref{eq:K_benchmarks_1} and \eqref{eq:K_benchmarks_2}.
                  }
\label{fig:coupling_gauge_bosons}
\end{figure}

The couplings to gauge bosons in \eqref{eq:L_1} are generated by the SM particles in one- and two-loop diagrams~\cite{Heeck:2019guh}.
The photon and gluon amplitudes contain the loop function $h(m_J^2/(4m_{\mathrm{SM}}^2))$, defined by
\begin{equation}
  h(x)= -\frac{1}{4 x}\Big(\log\!\big[1-2 x + 2\sqrt{x (x-1)}\big]\Big)^2 -1 \,.
\label{eq:h_1}
\end{equation}
Its asymptotic forms are
\begin{equation}
  h(x) \simeq \frac{x}{3}+\mathcal{O}(x^2)\ \ (x\to 0) \quad \mathrm{and} \quad h(x) \simeq -1 \ \ (x\to \infty)\,.
\label{eq:h_2}
\end{equation}
As a result, $|h(x)|$ is suppressed for small $x$, achieves its maximum at $x=1$, and rapidly approaches its asymptotic value $h(x)\to -1$ for $x\to\infty$.
This behavior explains the peak and trough structure of the mass dependence of gauge boson couplings shown in Fig.~\ref{fig:coupling_gauge_bosons}.

For qualitative estimates relevant to the phenomenology of the GeV-scale majoron, we present the functional dependence of the gauge boson couplings in the $m_J\to 0$ limit; in numerical analysis we use the full expressions~\cite{Heeck:2019guh}.
Given the behavior of the $h(x)$ function, the lightest SM particle in the loop provides the dominant contribution to the coupling, resulting in 
\begin{equation}
\label{eq:g_Jgg}
    g_{Jgg} \simeq \frac{\alpha_S \,m_J^2}{384\pi^3\, m_{u}^2\, v} \tr(K) 
     \,,
\end{equation}
\begin{align}
    g_{J\gamma\gamma} &\simeq \frac{\alpha\,m_J^2}{96\pi^3\,v}\Bigg[ \left(\frac{K_{ee}}{m_e^2}+\frac{K_{\mu\mu}}{m_\mu^2}+\frac{K_{\tau\tau}}{m_\tau^2}\right) -\frac{\tr K}{2 \,m_e^2} \Bigg]
\,,
\end{align}
For the on-shell $Z$ boson, the $JZ\gamma$ amplitude retains a nonzero limit as $m_J\to 0$, unlike the $J\gamma\gamma$ amplitude. 

The fermion sector plays a key phenomenological role.
Specializing \eqref{eq:L_1} to leptons and quarks, the relevant part of the Lagrangian is
\be
  {\cal L_{\mathrm{int.}}} &\supset 
        \frac{1}{2f} \, \partial_\mu J \,\, (\overline{e} \; \overline{\mu} \; \overline{\tau}) \,\gamma^\mu \left( C^{\ell}_V + C^{\ell}_A \, \gamma_5\right) \begin{pmatrix}
        e \\
        \mu \\
        \tau
    \end{pmatrix}\\
        &+ \frac{1}{2f} \, \partial_\mu J \,\, (\overline{u} \; \overline{c} \; \overline{t}) \,\gamma^\mu \left( C^u_V + C^u_A \, \gamma_5\right) \begin{pmatrix}
        u \\
        c \\
        t
    \end{pmatrix}\\
        &+\frac{1}{2f} \, \partial_\mu J \,\, (\overline{d} \; \overline{s} \; \overline{b}) \,\gamma^\mu \left( C^d_V + C^d_A \, \gamma_5\right) \begin{pmatrix}
        d \\
        s \\
        b
    \end{pmatrix}
\,.
\label{eq:L_2}
\ee
Matching the loop-induced couplings to leptons~\cite{Garcia-Cely:2017oco} to \eqref{eq:L_2} gives
\be
  C^{\ell}_V = \frac{-f \, K}{8\pi^2 v}\,, \,\,\,\,\,\,\, C^{\ell}_A = \frac{f}{8\pi^2 v} \left( K - \frac{\tr(K)}{2}\, \mathbb{1}_{3\times 3} \right)\,,
\label{eq:C_leptons}
\ee
where the relation for off-diagonal elements, $C_V^o=-C_A^o$, originates from the purely left-handed nature of the weak interaction, while the trace term affects only diagonal axial couplings.

For quarks, we obtain the analogous expressions,
\be
  C^{u}_V =C^{d}_V =0\,, \,\,\,\,\,\,\, C^{u}_A  = -C^{d}_A = \frac{f}{16\pi^2 v}  \tr(K) \, \mathbb{1}_{3\times 3} \,.
\label{eq:C_quarks}
\ee
Although $C^{\ell}$ and $C^{u/d}$ arise at the same order in perturbation theory and are parametrically similar, there are no FV quark couplings at one loop, and the two-loop result is additionally suppressed by a factor $\sim 1/(16\pi^2)$.
On the other hand, the values of the off-diagonal leptonic couplings, $C^{l_i \, l_j }_V$ and $C^{l_i \, l_j }_A$ for $i\neq j$, can be comparable to the flavor-conserving (FC) ones and dominate particular observables when the relevant diagonal entries are small or cancellations occur.
As a result, FV decays of heavy leptons, $\tau^\pm$ in particular, are an efficient production mechanism for $J$.

To obtain \eqref{eq:C_leptons} and \eqref{eq:C_quarks}, the equations of motion were used. 
For quarks, one needs to take into account the quark axial-current identity 
\be
  \partial_\mu(\bar q\gamma^\mu\gamma_5 q) = 2i m_q\,\bar q\gamma_5 q + \text{anomaly}\,,
\ee
where the anomaly terms are consistently taken into account when deriving the effective coupling to $G\tilde G$.
Moreover, when matching below $\Lambda_{\rm QCD}$ for light quarks, the anomaly terms must be combined with non-perturbative chiral physics.

Since the low-energy effective couplings in Eq.~\ref{eq:L_1} are not $SU(2)_L$ invariant after the EW symmetry breaking, the majoron can also be produced in helicity-suppressed decays of light pseudoscalars, such as pions and kaons~\cite{Altmannshofer:2022ckw}.
These channels are kinematically restricted to $m_J<m_P-m_\ell$ for $P^\pm\to\ell^\pm\nu J$, with $P=\pi,K$. We checked that these modes do not yield competitive limits compared to astrophysical probes or limits derived from $\tau^\pm$ decays.

The majoron couplings to nucleons are obtained by matching quark-level axial/pseudoscalar operators to nucleon matrix elements using the axial Ward identity (including the anomaly) together with nucleon axial charges $\Delta_q^N$. 
We define the nucleon couplings by
\be
\label{eq:L_JNN}
  \mathcal{L}_{JN\bar{N}}=iJ\sum_{N=p,n} C_{N\bar{N}}\bar N\gamma_5N
\,.
\ee
Next, we follow the standard treatment to evaluate this coupling by using the lattice inputs for $\Delta_q^N$. 
Using isospin symmetry and the axial charges, $\Delta_u^p = 0.847(37)$, $\Delta_d^p = -0.407(24)$, and $\Delta_s^p = -0.035(9)$ from Refs.~\cite{FlavourLatticeAveragingGroupFLAG:2024oxs,Alexandrou:2024ozj}, we obtain
\be
  C_{p\bar{p}}=\frac{-1.29(3)\, m_p \,\tr (K)}{16\pi^2 v}\,, \,\,\,\,\,\,\,\,\,\, C_{n\bar{n}}=\frac{1.22(3)\, m_n \,\tr (K)}{16\pi^2 v}
\,,
\label{eq:Gamma_JNN_1}
\ee
which is a percent-level correction to the result of~\cite{Heeck:2019guh}.

\textit{Benchmarks.}
We study representative effective coupling textures of GeV-scale majoron. 
Among several benchmarks, we investigate a bottom-up phenomenological parameterization introduced in~\cite{Bertuzzo:2022fcm}, which we call the ``anarchical'' benchmark.
In this case, the matrix $K$ defined by \eqref{eq:K_mat} has degenerate diagonal entries $K^d$ and degenerate off-diagonal entries $K^o$, parameterized by the ratio $R_J = |K^d/K^o|$,
\be
    K_{\rm anarch} = \begin{pmatrix} K^d & K^o & K^o \\ K^o & K^d & K^o \\ K^o & K^o & K^d \end{pmatrix}
    =  \frac{8\pi^2 \,v}{f}  \begin{pmatrix} R_J & 1 & 1 \\ 1 & R_J & 1 \\ 1 & 1 & R_J \end{pmatrix}
\,.
\label{eq:K_anarch}
\ee
The scale $f$ parameterizes the overall strength of the loop-induced couplings  and is in general distinct from the symmetry-breaking scale $f_J$. 
They are related by the relation $f\, yy^\dagger=16\pi^2 \,f_J \,\widehat K$, where $\widehat K$ denotes the dimensionless benchmark texture, $y$ is the  $3\times n_N$ neutrino Yukawa matrix.
An extended sterile-neutrino sector was already envisaged in~\cite{Bertuzzo:2022fcm} to motivate the $R_J=10$ benchmark, and we likewise allow realizations beyond three RHNs. 
For suitable $f_J/f$, the positive-definite benchmarks considered here admit perturbative realizations compatible with small light-neutrino masses, for example with six singlets and an approximate lepton-number symmetry, with all sterile masses generated by the same scalar $S$.

The eigenvalues of the $K_{\rm anarch}$ matrix are positive if and only if $K^d - K^o > 0$ and $K^d + 2K^o > 0$. 
For positive real $K^o$, this requires $R_J > 1$; for negative real $K^o$, one needs $R_J > 2$.
Similar considerations apply to the matrix 
\be
  \begin{pmatrix} K^d &\pm K^o & \pm K^o \\ \pm K^o & K^d & \pm K^o \\ \pm K^o & \pm K^o & K^d \end{pmatrix} \,,
\ee
for any choice of the signs as long as the resulting matrix is hermitian.
With the ansatz given by \eqref{eq:K_anarch}, the couplings to matter fields are as follows (see~\eqref{eq:L_1}):
\be
  \left( C^{\ell}_V \right)^{d} &= 2\left( C^{\ell}_A \right)^{d} = -R_J
  \,, \\
    \left( C^{\ell}_V\right)^{o} &= -\left( C^{\ell}_A\right)^{o} = -1
  \,, \,\,\,\,\, \\
    C^{u}_V &= C^{d}_V = \left( C^{u}_A \right)^{o} = \left( C^{d}_A \right)^{o} = 0
  \,, 
  \\
  \left( C^{u}_A \right)^{d} &= -\left( C^{d}_A \right)^{d} =\frac32 R_J 
\,.
\label{eq:C_structure_RJ}
\ee

Ref.~\cite{Bertuzzo:2022fcm} studied exclusively the anarchical benchmark with two values of the ratio: $R_J = 2/3$ and $R_J = 10$.
We note that only the latter benchmark leads to a positive-definite $K$.
For comparison, we show results for three such benchmarks: $R_J = 2/3$, $R_J = 1.01$, and $R_J = 10$. The choice $R_J = 1.01$ lies close to the positive-definiteness boundary.

Secondly, we investigate single-flavor textures,
\be
  K_{\mu\tau}  = K^d\begin{pmatrix} 1 & 0 & 0 \\ 0 & 1 & \kappa_{\mu\tau} \\ 0 & \kappa_{\mu\tau}^* & 1 \end{pmatrix} 
\,,
\label{eq:K_benchmarks_1}
\ee
with $|K^d|=8\pi^2 v/f$, and where the $\kappa_{\alpha\beta}$ parameters are bounded by the requirement of positive-definiteness of the $K$ matrix: $|\kappa_{\alpha\beta}| < 1$.
We study cases $\kappa_{\mu\tau}=0.01$ and $\kappa_{\mu\tau}=0.5$, which correspond to hierarchical and anarchical regimes between the diagonal and off-diagonal elements, respectively.
The couplings to the SM fermions read as follows:
\be
  \left( C^{\ell}_V \right)^{d} = 2\left( C^{\ell}_A \right)^{d} = -1
\,, 
\\
  \left( C^{\ell}_V\right)^{\mu\tau} = -\left( C^{\ell}_A\right)^{\mu\tau} = -\kappa_{\mu\tau}
\,,
\\
  C^{u}_V = C^{d}_V = \left( C^{u}_A \right)^{o} = \left( C^{d}_A \right)^{o} = 0
\,, 
\\
  \left( C^{u}_A \right)^{d} = -\left( C^{d}_A \right)^{d} =\frac32
\,.
\label{eq:C_structure_single_flavor_textures}
\ee

Finally, we study complex-valued benchmarks with the following parameterization:
\be
  K^{\mathrm{CP}}  = K^d\begin{pmatrix} 1 & i\,\kappa &  i\,\kappa \\ -i\,\kappa & 1 & i\,\kappa \\ -i\,\kappa & -i\,\kappa & 1 \end{pmatrix} 
\,,
\label{eq:K_benchmarks_2}
\ee
where $\kappa < 1/\sqrt{3}$ is a positive real parameter, as required by positive-definiteness of $K^{\mathrm{CP}}$. 
We study cases $\kappa = 0.01$ and $\kappa = 0.577$, where the latter is chosen to nearly saturate the positive-definiteness bound $\kappa < 1/\sqrt{3} \approx 0.5774$.
The structure of the couplings to the SM fermions is 
\be
  \left( C^{\ell}_V \right)^{d} = 2\left( C^{\ell}_A \right)^{d} = -1
\,, 
\\
  \left( C^{\ell}_V\right)^{\alpha\beta} = -\left( C^{\ell}_A\right)^{\alpha\beta} = -i\,\kappa
\,,
\\
  C^{u}_V = C^{d}_V = \left( C^{u}_A \right)^{o} = \left( C^{d}_A \right)^{o} = 0
\,, 
\\
  \left( C^{u}_A \right)^{d} = -\left( C^{d}_A \right)^{d} =\frac32
\,.
\label{eq:C_CP}
\ee
Here $(\alpha,\beta)=(e,\mu),(e,\tau),(\mu,\tau)$, and the lower off-diagonal entries are fixed by Hermiticity.

\section{Majoron production and decay modes}
\label{sec:prod_decays}

\paragraph{$J$ production modes.} 

The leading majoron production modes are LFV $\tau^\pm$ decays, $\tau^\pm\to \mu^\pm J$ and $\tau^\pm\to e^\pm J$.
The corresponding kinematic endpoints are $m_J<m_\tau-m_\mu$ and $m_J<m_\tau-m_e$, respectively.
The decay widths for FV decays of a fermion $\psi_i$ into  $\psi_j$ and $J$ are given in Ref.~\cite{Heeck:2017xmg} for real coefficients $C^{ij}_{A}$ and $C^{ij}_{V}$. 
Below, we write expressions valid for the general case, which is needed, \eg, for our complex benchmarks: 
\begin{equation}
  \small
  \begin{aligned}
    \Gamma(\psi_i \to \psi_j\,J) ={}& \frac{m_i^3}{64\pi f^2} \sqrt{\left(1-r_J^2\right)^2+r_j^4 -2r_j^2\left(1+r_J^2\right)} \\[2pt] &\times\Big\{ \big[|C_A^{ij}|^2+|C_V^{ij}|^2\big](1-r_j^2)^2 \\[-1pt] &\qquad -\big[|C_V^{ij}|^2(1-r_j)^2 +|C_A^{ij}|^2(1+r_j)^2\big]r_J^2 \Big\} \\[2pt] \simeq{}& \frac{m_i^3}{64\pi f^2} \big[|C_A^{ij}|^2+|C_V^{ij}|^2\big]
  \,.
  \end{aligned}
\end{equation}
where $r_{j,J} = m_{j,J}/m_i$ and the last relation holds for $r_{j,J} \ll 1$.

Subdominant production modes are three-body meson decays, three- or four-body lepton decays, and proton bremsstrahlung~\cite{Bertuzzo:2022fcm,Ema:2025bww}.
We verified that these channels yield negligible event rates compared to the $\tau^\pm$ decays for all the benchmarks considered, and we do not discuss them further.

At beam dump experiments and LHC far detectors, the $\tau^\pm$ spectrum is generated dominantly by leptonic decays of forward-produced charm mesons, primarly $D_s^\pm\to\tau^\pm\nu_\tau$ and, subdominantly, $D^\pm\to\tau^\pm\nu_\tau$. 
We simulated forward charm production using \textsc{Pythia\,8} with the Monash tune~\cite{Sjostrand:2014zea,
Skands:2014pea} by using \texttt{SoftQCD:all} and disabling \texttt{HardQCD} to avoid double counting.
For the LHC $\tau^\pm$ spectrum we also used the dedicated MC generator \textsc{SIBYLL}~\cite{Ahn:2009wx}, finding results consistent with those obtained using \textsc{Pythia\,8}.
In the simulation, we decay $D_s^\pm$ and $D^\pm$ to $\tau^\pm$ and weight each event by the corresponding branching fraction.
Since the SM leptonic branching ratio is helicity-suppressed,
\begin{equation}
  \mathrm{BR}(D_s^\pm \to \ell^\pm \nu_\ell)=\frac{G_F^2}{8\pi}\,\tau_{D_s}\,f_{D_s}^2\,|V_{cs}|^2\, m_{D_s}\,m_\ell^2\left(1-\frac{m_\ell^2}{m_{D_s}^2}\right)^2,
\end{equation}
implying $\mathrm{BR}(D_s^\pm\to \mu^\pm\nu)/\mathrm{BR}(D^\pm_s\to \tau^\pm\nu)\simeq 0.10$ and a negligible electron mode.

\vspace{10pt}

\paragraph{$J$ decays.} 
We discuss the available decays of the majoron originating from its couplings given by \eqref{eq:L_1} and \eqref{eq:L_2}.

The loop-induced coupling to photons gives the diphoton decay width
\be
  \Gamma(J\to \gamma\gamma) = \frac{|g_{J\gamma\gamma}|^2\,m_J^3}{64\pi}
\,.
\ee

\begin{figure}[tb]
      \centering
          \includegraphics[scale=0.4]{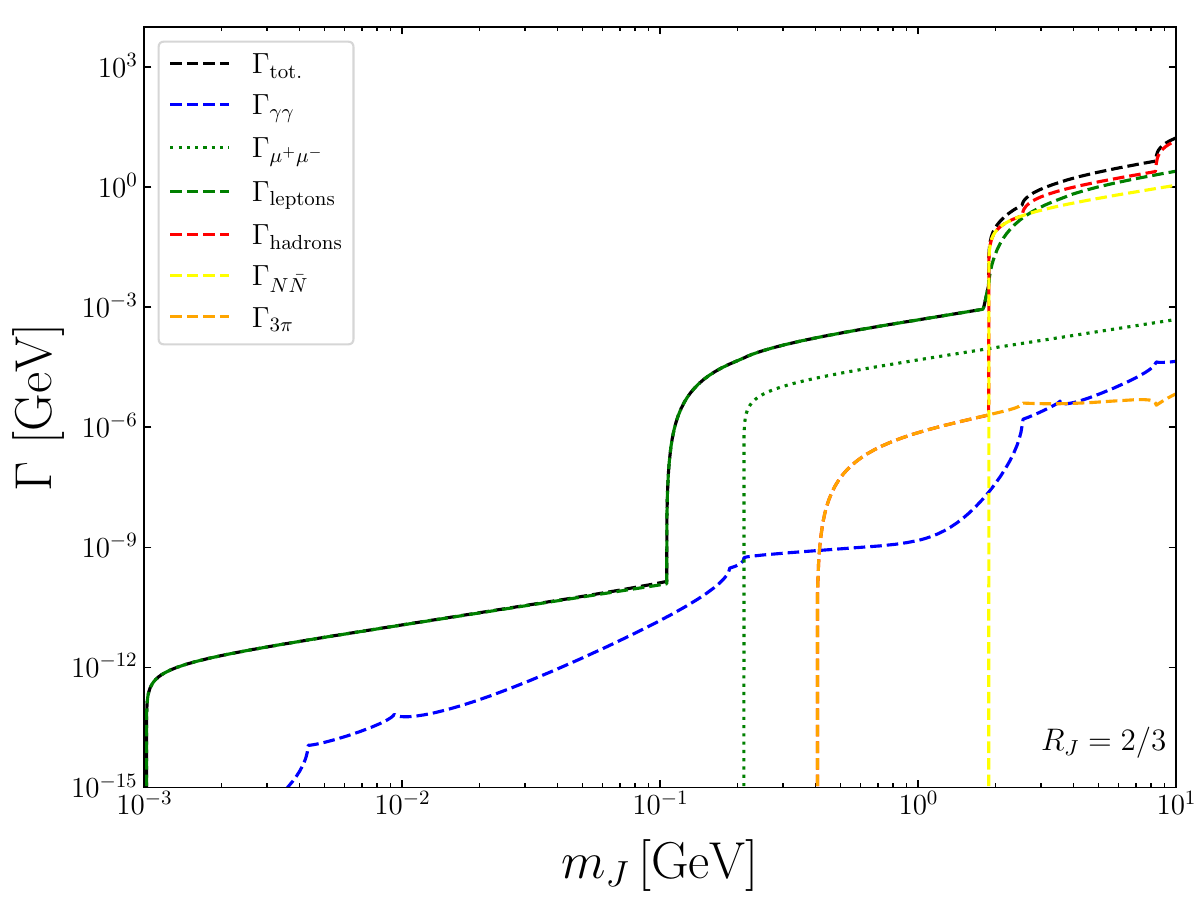}
          
          \hspace{0.1cm}

          \includegraphics[scale=0.4]{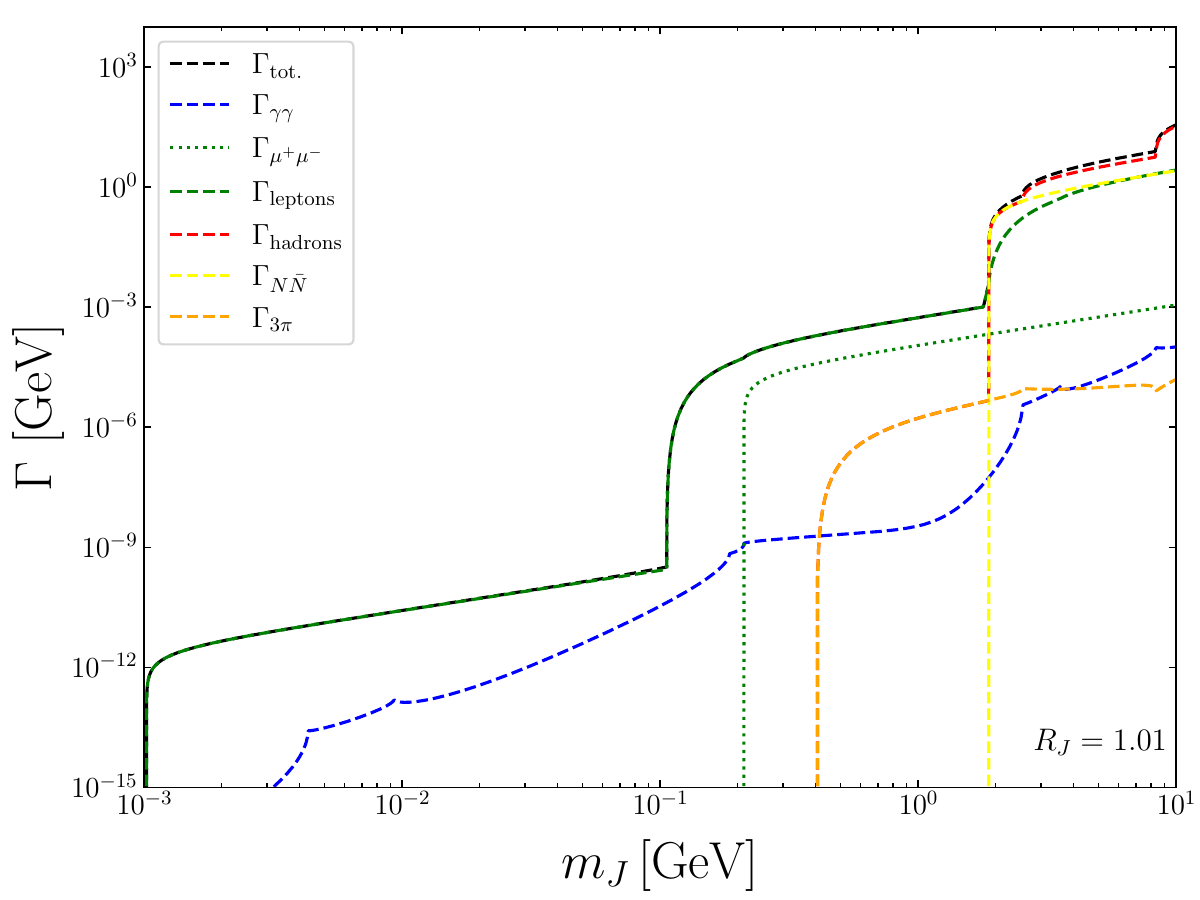}
          
          \hspace{0.1cm}
          
          \includegraphics[scale=0.4]{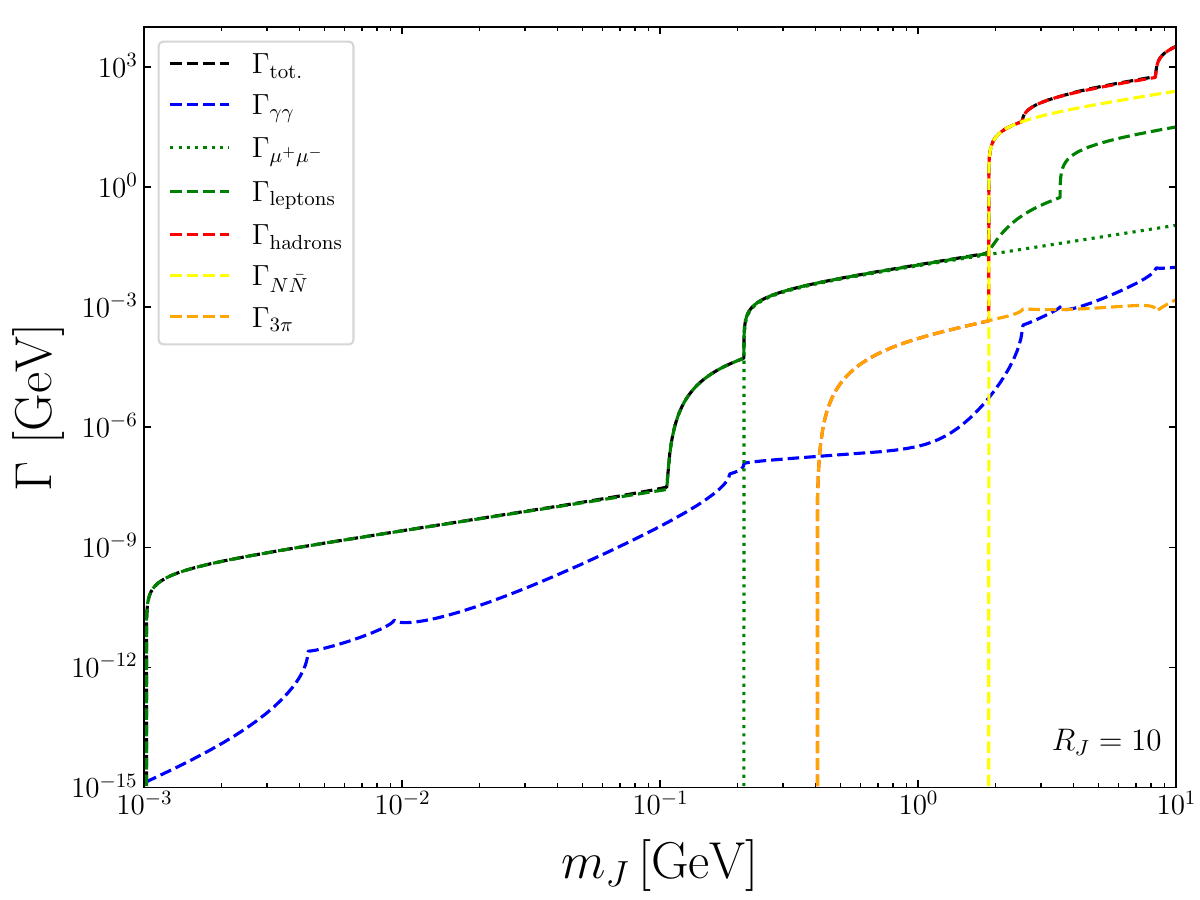}
          \caption{
                  Majoron partial decay widths for $K=K_{\rm anarch}$ benchmarks: $R_J=2/3$ (top), $R_J=1.01$ (center), and $R_J=10$ (bottom).
                  We fixed $f=1$ GeV, while the general case is obtained by rescaling by $(1\,\mathrm{GeV}/f)^2$.
                  }
\label{fig:decay_widths}
\end{figure}
\begin{figure}[tb]
      \centering
          \includegraphics[scale=0.4]{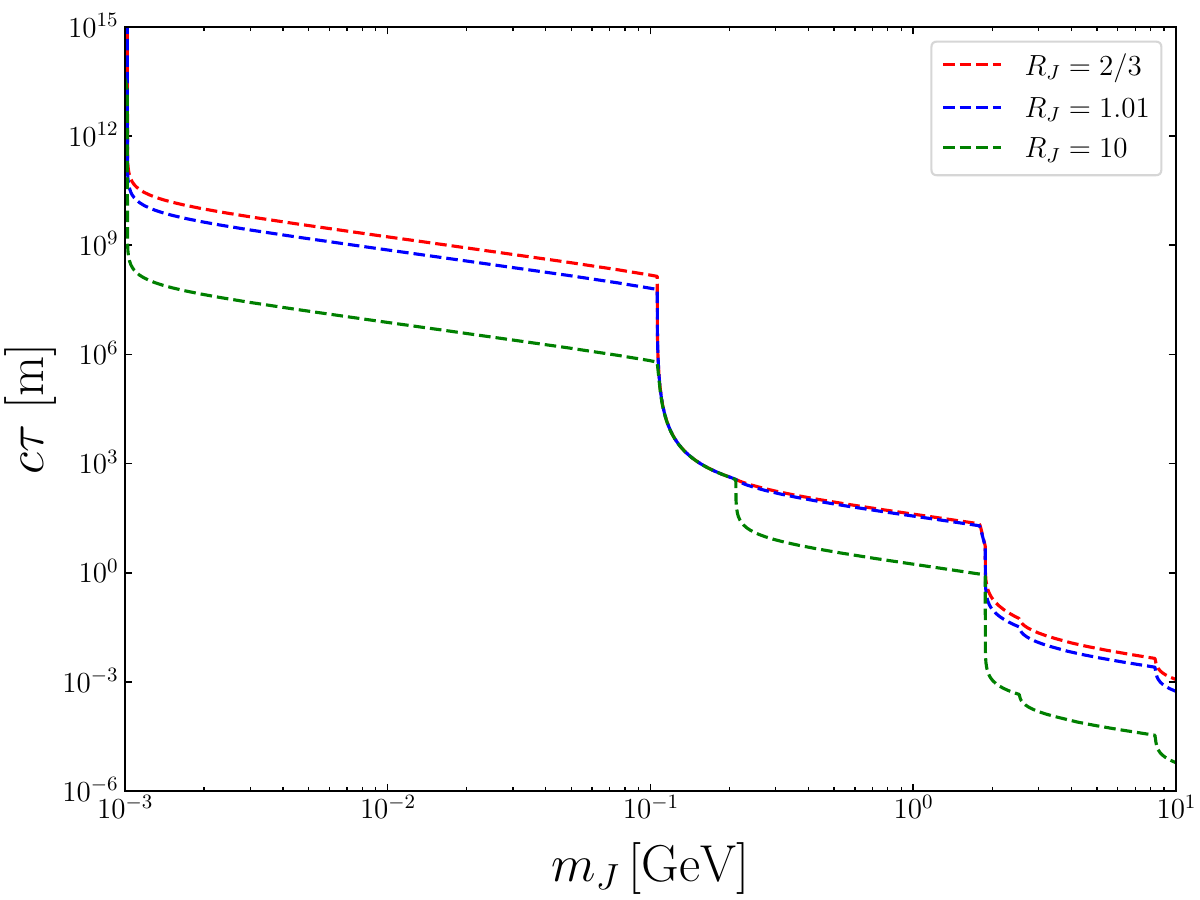}
          \caption{
                  Majoron proper decay length $c\tau$ for the anarchical benchmarks: $R_J=2/3$ (red), $R_J=1.01$ (blue), and $R_J=10$ (green).
                  We fixed $f=10^7$ GeV, which corresponds to a typical value that can be probed at IF experiments (see Sec.~\ref{sec:results}).
                  The general case is obtained by rescaling by $(f/10^7\,\mathrm{GeV})^2$.
                  }
\label{fig:ctau}
\end{figure}

Above the electron pair threshold, the majoron can decay into charged SM fermions. 
For charged leptons, the FC decay widths are given by
\begin{align}
  &\Gamma (J\to \psi_i \, \bar\psi_i) = \frac{m_i^2\,m_J}{8\pi\,f^2} \, \left|C^{ii}_{A}\right|^2\, \sqrt{1-\frac{4m_i^2}{m_J^2}}
\,.
\end{align}
The FV ones are described by 
\begin{equation}
  \small
  \begin{aligned}
    \label{eq:J_FV}
    \Gamma(J\to\psi_i\bar\psi_j+\bar\psi_i\psi_j) ={}& \frac{\sqrt{(m_i^2+m_j^2-m_J^2)^2-4m_i^2m_j^2}} {16\pi f^2 m_J^3} \\[2pt] &\times\Big\{ m_J^2\Big[|C_A^{ij}|^2(m_i+m_j)^2 \\[-1pt] &\qquad\quad +|C_V^{ij}|^2(m_i-m_j)^2\Big] \\[-1pt] &\qquad -\big[|C_A^{ij}|^2+|C_V^{ij}|^2\big] (m_i^2-m_j^2)^2 \Big\} \\[2pt] \simeq{}& \frac{m_i^2m_J}{16\pi f^2} \big[|C_A^{ij}|^2+|C_V^{ij}|^2\big] \left(1-\frac{m_i^2}{m_J^2}\right)^2
\,,
\end{aligned}
\end{equation}
where the last expression holds for $m_j\ll m_i$. 
Eq.~\ref{eq:J_FV} generalizes the approximate result obtained in Ref.~\cite{Heeck:2017xmg} for real $C_A^{ij}$ and $C_V^{ij}$ to complex couplings.
Therefore, for non-zero FV couplings and when kinematically allowed, the charged-lepton final states are $e^+ e^-$, $\mu^{\pm} e^{\mp}$, $\mu^+ \mu^-$, $e^{\pm} \tau^{\mp}$, $\mu^{\pm} \tau^{\mp}$, and $\tau^- \tau^+$.

For $3m_\pi < m_J < 2m_N$, where $m_N$ is the nucleon mass, decays into light hadrons become kinematically accessible. Since for the majoron the diagonal axial quark and lepton couplings are of similar strength (see \eqref{eq:C_quarks}), the hadronic decays may become important for $m_J\simeq 1\,\gev$. 
However, for the benchmarks considered, the calculated three-pion widths are subdominant to the leptonic widths. 
We used the public code ALPaca~\cite{Alda:2025nsz} to verify the suppression of the hadronic rates at representative benchmark points in the relevant mass range.
For a qualitative understanding of the hadronic decays, the $J \to 3\pi$ mode is especially relevant, since it is the leading hadronic mode for an ALP with $m\lesssim 1\,$GeV~\cite{Aloni:2018vki}.
Following Ref.~\cite{Bauer:2017ris}, the three-pion widths are
\begin{align}
\label{eq:J_to_pi0_pi0_pi0}
  &\Gamma(J\to\pi^0\pi^0\pi^0) =\left|\frac{C_A^{uu}-C_A^{dd}}{32\pi^2 f}\right|^2 \frac{\pi m_Jm_{\pi^0}^4}{6f_\pi^2} \frac{2}{(1-r_0)^2}\nonumber\\ &\quad\times\int_{4r_0}^{(1-\sqrt{r_0})^2} dz\,\sqrt{1-\frac{4r_0}{z}}\, \lambda^{1/2}(1,z,r_0)
\,,
\end{align}
\begin{align}
\label{eq:J_to_pi0_piplus_piminus}
  &\Gamma(J\to\pi^0\pi^+\pi^-) =\left|\frac{C_A^{uu}-C_A^{dd}}{32\pi^2 f}\right|^2 \frac{\pi m_Jm_{\pi^+}^4}{6f_\pi^2} \frac{12}{(1-r_+)^2}\nonumber\\ &\quad\times\int_{4r_+}^{(1-\sqrt{r_+})^2} dz\,\sqrt{1-\frac{4r_+}{z}}\, (z-r_+)^2\lambda^{1/2}(1,z,r_+)
\,,
\end{align}
where $r_a= \frac{m_{\pi^a}^{2}}{m_J^{2}}$, $a\in\{0,+\}$, $\lambda(x,y,z)=x^2+y^2+z^2-2xy-2xz-2yz$, and $f_\pi\simeq130\,\mathrm{MeV}$.
We numerically implemented \eqref{eq:J_to_pi0_pi0_pi0} and~\eqref{eq:J_to_pi0_piplus_piminus}, finding that $J \to 3\pi$ decays are subdominant for the benchmarks studied, as shown in Fig.~\ref{fig:decay_widths}.

For $2m_N < m_J<2\,\mathrm{GeV}$, decays of $J$ into a pair of nucleons become relevant. 
From \eqref{eq:L_JNN}, the resulting decay widths are then
\be
  \Gamma_{J\to N\bar{N}} = |C_{N\bar{N}}|^2 \, \frac{m_J\sqrt{1 - 4m_N^2/m_J^2}}{8\pi} 
\,.
\label{eq:Gamma_JNN}
\ee

For $m_J\gg 2\,\gev$, inclusive hadronic decay widths can be computed using perturbative QCD and invoking quark-hadron duality~\cite{Bauer:2017ris,Shifman:2000jv,Wang:2025ncc},
\begin{align}
\label{eq:Gamma_J_had}
	\Gamma(J\to \textrm{hadrons}) &\simeq 32\pi\,\alpha_s^2(m_J)\,m_J^3 \nonumber \\
  &\times  \left[1+\frac{83}{4}\frac{\alpha_s(m_J)}{\pi}\right]  \left| \sum_{q=u,d,s} \frac{C_A^{qq}}{32 \pi^2 f} \right|^2
\,.
\end{align}

\section{Phenomenology: signatures and experimental setup}
\label{sec:phenomenology}
The observable signal consists of high-energy SM particles produced in majoron decays. We consider final states containing either a pair of photons or a pair of charged leptons, including both FC and FV configurations.

The decay probability for these processes within a detector region of length $\Delta = L_{\mathrm{max}} - L_{\mathrm{min}}$ is given by
\be
      p(d) = e^{-L_{\mathrm{min}}/d} - e^{-L_{\mathrm{max}}/d}
\,,
\label{eq:prob_decay}
\ee
with $d = \gamma \beta c\tau $ representing the boosted decay length of the long-lived particle (LLP) in the laboratory frame, and $L_{\mathrm{min}}$ denoting the separation between the LLP production vertex and the detector's upstream boundary.

Following LLP production, the expected number of observable events exhibiting the characteristic LLP decay signature within the fiducial volume is
\be
  N = \int \int dE\, d\theta \,\frac{d^2 N}{dE d\theta}\,\, p(E, \theta)\, q_{\text{acc.}}(E, \theta)
\,,
\label{eq:NoE}
\ee
where $\frac{d^2 N}{dE d\theta}$ represents the differential LLP distribution in energy $E$ and polar angle $\theta$ with respect to the beam axis.
The factor $p(E, \theta)$ encodes the decay probability within the detector acceptance, and $q_{\text{acc.}}(E,\theta)$ captures the geometric and kinematic acceptance after applying selection criteria.

At lepton colliders running at a center-of-mass energy equal to the $Z$ boson mass, the number of events resulting from direct $Z$ boson decays is
\be
  N = N_Z \times \mathrm{BR}(Z\to J \gamma) \times p_{J\to \mathrm{SM} + \mathrm{SM}}(E, \theta)
\,,
\label{eq:NoE_FCC}
\ee
where we use $N_Z=2.5\times10^{12}$ for FCC-ee~\cite{Chrzaszcz:2020emg} and $N_Z=4\times 10^{12}$ for CEPC~\cite{Ai:2025CEPC}.
The factor $\mathrm{BR}(Z\to J \gamma)$ is obtained from the decay width of the $Z$ boson into a photon and a majoron, which reads as follows:
\be
  \Gamma_{Z \to J \gamma} = \frac{|g_{J\gamma Z}|^2 \, m_Z^3 \,(1 - m_J^2/m_Z^2)^3}{96 \pi} 
\,.
\label{eq:Gamma_Z_gammaJ}
\ee
While Ref.~\cite{Bauer:2017ris}, studying axion-like particle decays into $\gamma\gamma$ at FCC-ee under the assumption of negligible SM background, SM triphoton processes can be non-negligible~\cite{dEnterria:2023wjq}. 
Moreover, the decay width of $J\to\gamma\gamma$ is suppressed compared to leptonic decays due to the behavior of the $g_{J\gamma\gamma}$ coupling.
As a result, we checked that for production through $Z\to J\gamma$, even the optimistic zero-background assumption gives an FCC-ee reach of roughly $1/f\sim10^{-3}$--$10^{-2}\,\mathrm{GeV}^{-1}$ in the diphoton channel.
Including leptonic majoron decays leads to a reach of order $1/f\sim10^{-4}$--$10^{-3}\,\mathrm{GeV}^{-1}$, which is still weaker than beam dump and LFV searches in the parameter region of interest. We therefore omit the $Z\to J\gamma$ contribution from Figs.~\ref{fig:sensitivity_plot_1}--\ref{fig:sensitivity_plot_2}.
However, $Z\to\tau^+\tau^-$ decays provide an additional source of majorons through the LFV decays $\tau^\pm\to\ell^\pm J$, with $\ell=e,\mu$.
We estimate the resulting sensitivity to displaced dimuon decays using
\be
  N_{\mu^+\mu^-}&=N_{\tau^\pm}\,\mathrm{BR}(J\to\mu^+\mu^-) \\
  &\times \sum_{\ell=e,\mu}\mathrm{BR}(\tau^\pm\to\ell^\pm J)\,
  \epsilon_\ell(m_J,c\tau_J)
\,,
\label{eq:Zfactory_tau_yield}
\ee
where $N_{\tau^\pm}=2N_Z\,\mathrm{BR}(Z\to\tau^+\tau^-)$ counts individual tau leptons of both charges.
The adopted luminosities give $N_{\tau^\pm}=1.68\times10^{11}$ for FCC-ee and $2.69\times10^{11}$ for CEPC.
The efficiency $\epsilon_\ell$ includes the majoron decay probability, geometrical acceptance, reconstructed mass and energy requirements, common-versus-separated vertex fits, and an opposite-$\tau\to\pi\nu$ tag.
For both colliders, we evaluate it using a simplified Gaussian track-response model and signal efficiencies are evaluated at masses between $0.1$ and $1.77\,\mathrm{GeV}$.
Following~\cite{Ardu:2026vsr}, we require a reconstructed majoron flight distance between $1\,\mathrm{mm}$ and $1\,\mathrm{m}$ from the tau decay vertex, and the track response is assumed within $R<300\,\mathrm{mm}$ and $|z|<500\,\mathrm{mm}$, with unit tracking and particle-identification efficiencies before selection. 

As the main source of SM background, we simulated
\be
  \tau^-\to e^-\mu^+\mu^-\nu_\tau\bar\nu_e\,,\qquad
  \tau^-\to\mu^-\mu^+\mu^-\nu_\tau\bar\nu_\mu
\,,
\ee
and their charge conjugates with \textsc{MadGraph5\_aMC@NLO}.
No events passed the selection in samples of $10^5$ events per channel at the relevant mass windows; the corresponding $95\%$ upper bounds are below $0.3$ selected events at FCC-ee and $0.5$ at CEPC.
To allow for possible residual sources, \eg, hadronic tau decays with misidentification or decay in flight, we assume a residual background of $b=10$  events at FCC-ee, corresponding to $b=16$ at CEPC.
Including the simulated leptonic backgrounds, we show in Figs.~\ref{fig:sensitivity_plot_1}--\ref{fig:sensitivity_plot_2} contours requiring approximately $8.1$ and $9.7$ selected signal events at FCC-ee and CEPC, respectively.
While a complete detector and background study is beyond the scope of this work, we find that assuming $b=1$ or $b=100$ events at FCC-ee, does not significantly change these contours.
Hence, $Z$ boson factories can provide complementary but subleading sensitivity to the GeV-scale majoron.

Finally, we discuss hadron colliders.
The majoron has been studied at the LHC through Drell-Yan production, vector-boson fusion, and processes involving sterile neutrinos~\cite{majoron_LHC}.
However, the resulting limits are weak due to the suppressed couplings and small production cross sections, so we do not discuss them further.

\textit{Experimental setup.} 
In Table~\ref{tab:experiments}, we list the key parameters of the detectors considered, together with relevant references.
We consider both running (FASER) and proposed (MATHUSLA) LHC far detectors, as well as dedicated past and future beam dump experiments.
We investigate two FASER configurations for the High-Luminosity LHC: an upgraded version of the current detector, called FASER2, and the larger FPF FASER2 detector, proposed for the dedicated Forward Physics Facility~\cite{Anchordoqui:2021ghd}.
\begin{table}[t]
  \centering
  \caption{
    Technical specifications, and energy thresholds used in the MC simulations, for the detectors considered in this study.
    The second column gives the energy of the proton beam or center-of-mass energy of $pp$ collisions for the LHC-based experiments.
    The third column gives the exposure as either the number of protons on target or the integrated luminosity.
    Geometry lists transverse side lengths or radius $r$ in meters, except for MATHUSLA's three-dimensional size. This highly off-axis detector probes the high-$p_T$ regime complementary to forward detectors, and we follow the updated 40-meter modular design~\cite{MATHUSLA:2025eth}.
  }
  \label{tab:experiments}
  \footnotesize
  \setlength{\tabcolsep}{1.5pt}
  \renewcommand{\arraystretch}{1.12}
  \begin{ruledtabular}
    \begin{tabular}{lcccccc}
      \makecell[l]{Experiment\\and ref.}
      & \makecell{Energy\\(GeV)}
      & Exposure
      & \makecell{Geometry\\(m)}
      & \makecell{$L_{\min}$\\(m)}
      & \makecell{$\Delta$\\(m)}
      & \makecell{$E_{\min}$\\(GeV)} \\
      \hline
      \makecell[l]{ArgoNeuT\\target\\\cite{Anderson:2012vc,Bertuzzo:2022fcm}}
      & 120 & $1.09\times10^{20}$ & $0.4\times0.47$ & 1033 & 1.3 & 10 \\[4pt]
      \makecell[l]{ArgoNeuT\\absorber\\\cite{Anderson:2012vc,Bertuzzo:2022fcm}}
      & 120 & $1.63\times10^{19}$ & $0.4\times0.47$ & 318 & 1.3 & 10 \\[4pt]
      \makecell[l]{BEBC\\\cite{WA66:1985mfx,Barouki:2022bkt}}
      & 400 & $2.7\times10^{18}$ & $3.6\times2.5$ & 404 & 1.85 & 0.5 \\[4pt]
      \makecell[l]{CHARM\\\cite{CHARM:1985anb,Dobrich:2019dxc}}
      & 400 & $2.4\times10^{18}$ & $3\times3$ & 480 & 35 & 0.5 \\[4pt]
      \makecell[l]{DUNE\\\cite{DUNE:2021tad}}
      & 120 & $1.1\times10^{22}$ & $5\times5$ & 574 & 5.88 & 0.5 \\[4pt]
      \makecell[l]{FASER2\\\cite{Feng:2018pew}}
      & 13600 & $3\,\mathrm{ab}^{-1}$ & $r=1$ & 650 & 10 & 50 \\[4pt]
      \makecell[l]{FPF FASER2\\\cite{Feng:2022inv,Jodlowski:2020vhr}}
      & 13600 & $3\,\mathrm{ab}^{-1}$ & $r=1$ & 620 & 25 & 100 \\[4pt]
      \makecell[l]{MATHUSLA\\\cite{MATHUSLA:2025eth,Jodlowski:2019ycu}}
      & 13600 & $3\,\mathrm{ab}^{-1}$
      & \makecell{$40\times40$\\$\times11$} & $\sim100$ & 40 & 1 \\[4pt]
      \makecell[l]{NA62\\\cite{Dobrich:2019dxc}}
      & 400 & $1.0\times10^{18}$ & $r=1.13$ & 81 & 135 & 3 \\[4pt]
      \makecell[l]{NuCal\\\cite{Blumlein:2013cua,Dobrich:2019dxc}}
      & 69 & $1.7\times10^{18}$ & $r=1.3$ & 64 & 23 & 10 \\[4pt]
      \makecell[l]{SHiP\\\cite{Dobrich:2019dxc}}
      & 400 & $6.0\times10^{20}$ & $2.5\times5.5$ & 33.7 & 50 & 0.5 \\
    \end{tabular}
  \end{ruledtabular}
\end{table}

\section{Results}
\label{sec:results}

\begin{figure*}[tb]
      \centering
          \includegraphics[scale=0.282]{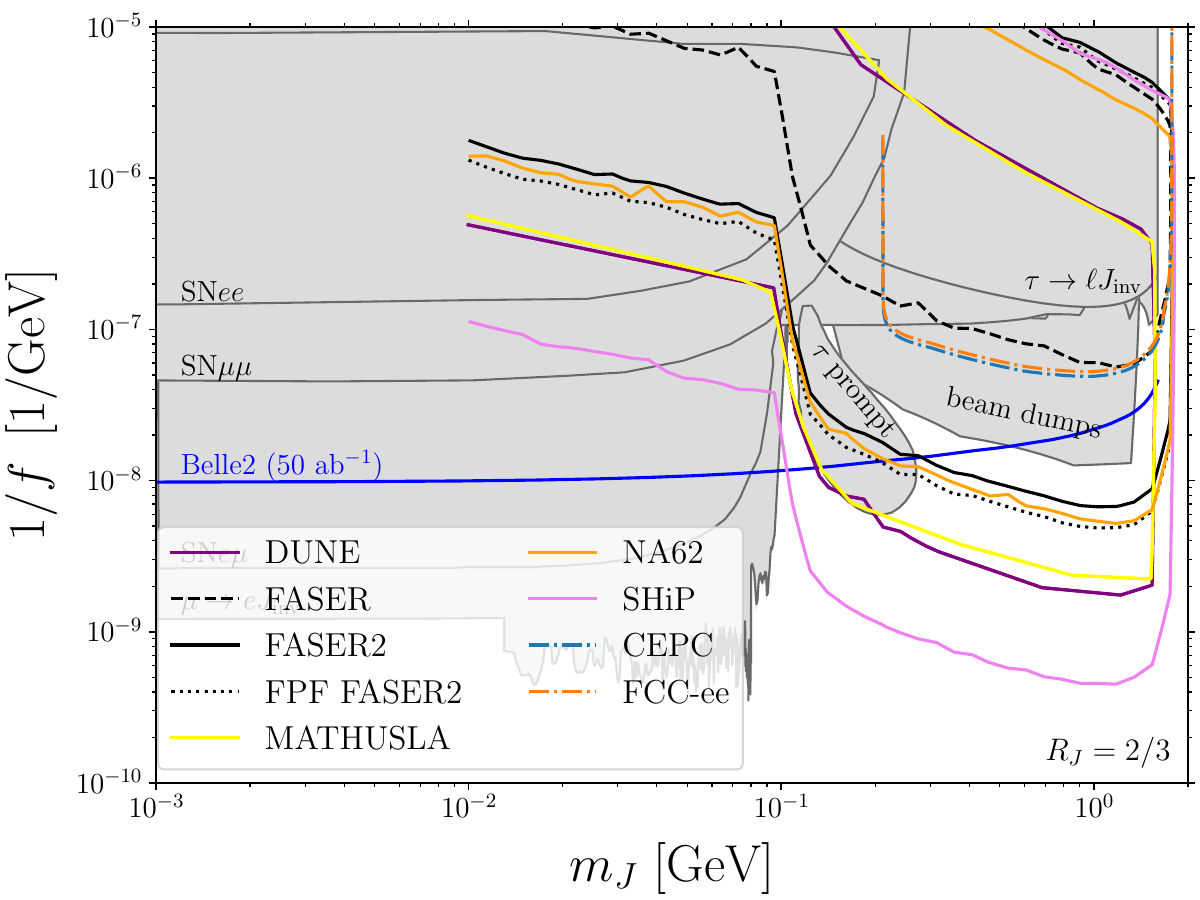}
          \hspace{0.01cm}
          \includegraphics[scale=0.288]{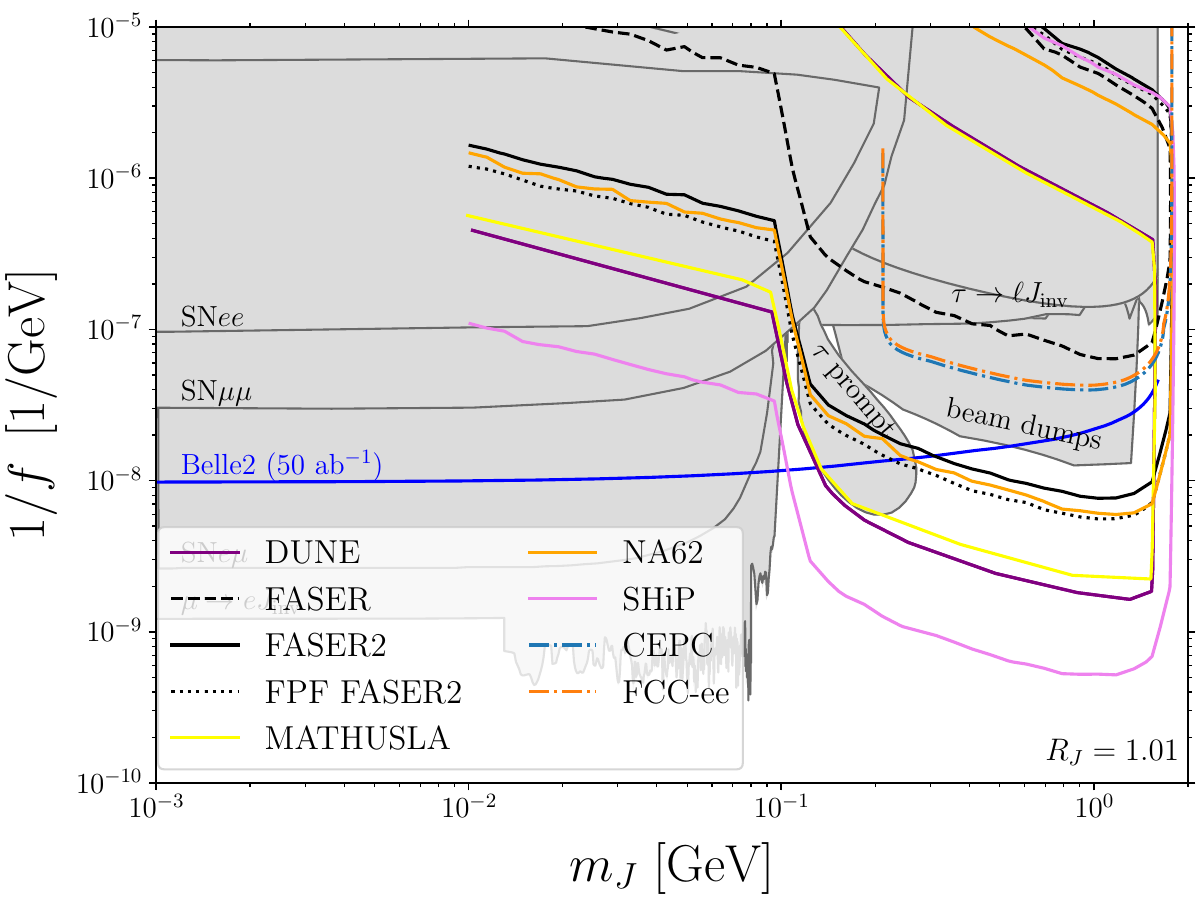}
          \hspace{0.01cm}
          \includegraphics[scale=0.288]{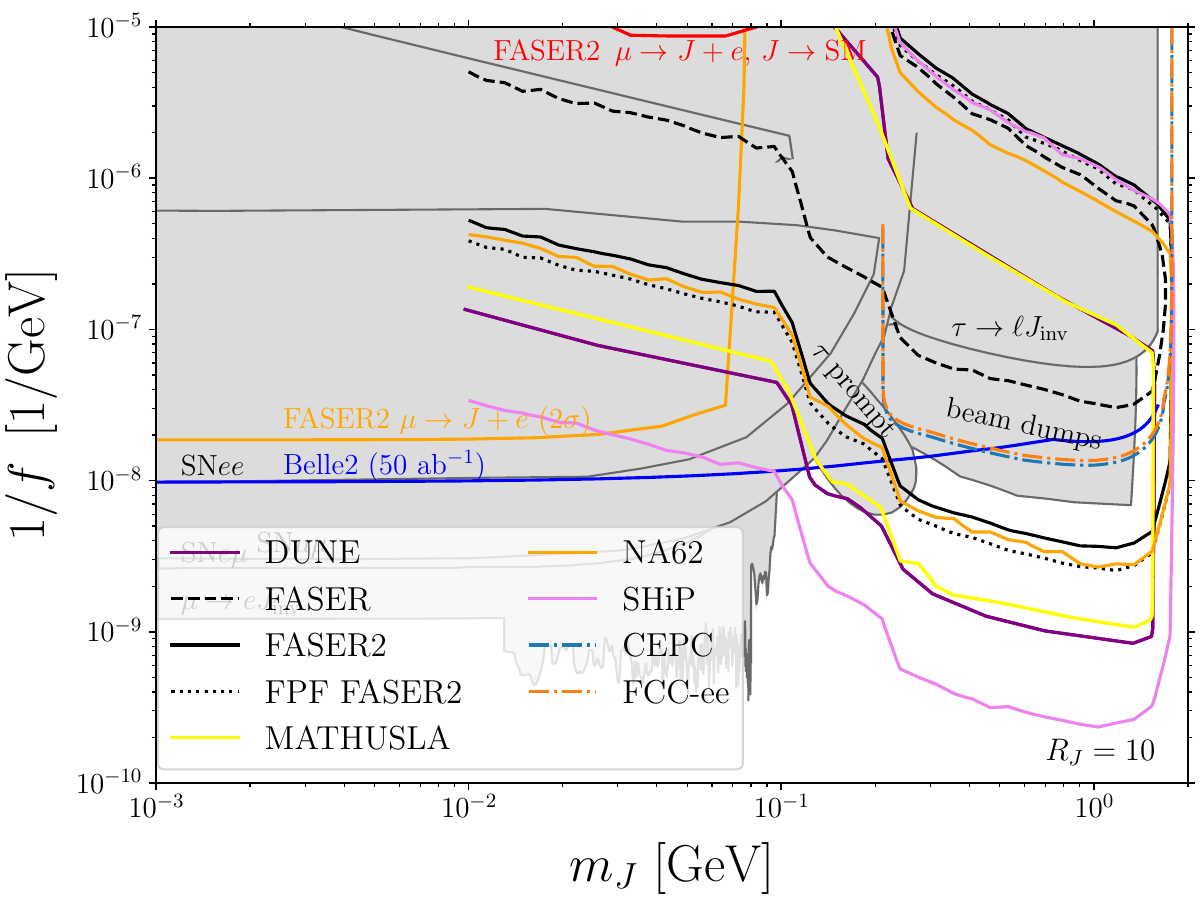}
          \caption{
                  Results for the ``anarchical'' benchmark (see~\eqref{eq:K_anarch}), shown in the $m_J$ vs.\ $1/f$ plane. 
                  For beam dump experiments and LHC far detectors, majorons are produced primarily via $\tau^\pm\to\ell^\pm J$ from forward $D_s/D\to\tau\nu_\tau$ decays, and both FC and FV charged-lepton decay channels are included.
                  The dash-dotted CEPC and FCC-ee curves instead use $Z\to\tau^+\tau^-$ followed by $\tau^\pm\to\ell^\pm J$ and $J\to\mu^+\mu^-$.
                  The gray-shaded region denotes the combined exclusion from past beam dump experiments (BEBC, CHARM, NuCal). 
                  Hatched and shaded regions show existing astrophysical constraints from supernova cooling (SN-$ee$, SN-$\mu\mu$, SN-$e\mu$) and terrestrial LFV searches ($\mu\to e\,J_{\rm inv.}$, $\tau\to\mu\,J_{\rm inv.}$, and $\tau$ lifetime). 
                  Colored curves show projected reaches for the listed future facilities (see legend).
                  The hierarchy between off-diagonal and diagonal elements is controlled by three values of $R_J$: $R_J=2/3$ (left), $R_J=1.01$ (center), and $R_J=10$ (right). 
                  In the right panel, we additionally show the reach from $\mu^\pm\to e^\pm J$ production using the high-energy muon flux traversing FASER2 (orange shading) and from prompt majoron decays following muon decay (red shading); these limits are weaker for the other benchmarks and thus are omitted there. 
                  Note that the $R_J=2/3$ benchmark does not yield a positive-semi-definite $K$ matrix and is shown for comparison only (see text).
                  }
\label{fig:sensitivity_plot_1}
\end{figure*}
\begin{figure*}[tb]
      \centering
          \includegraphics[scale=0.43]{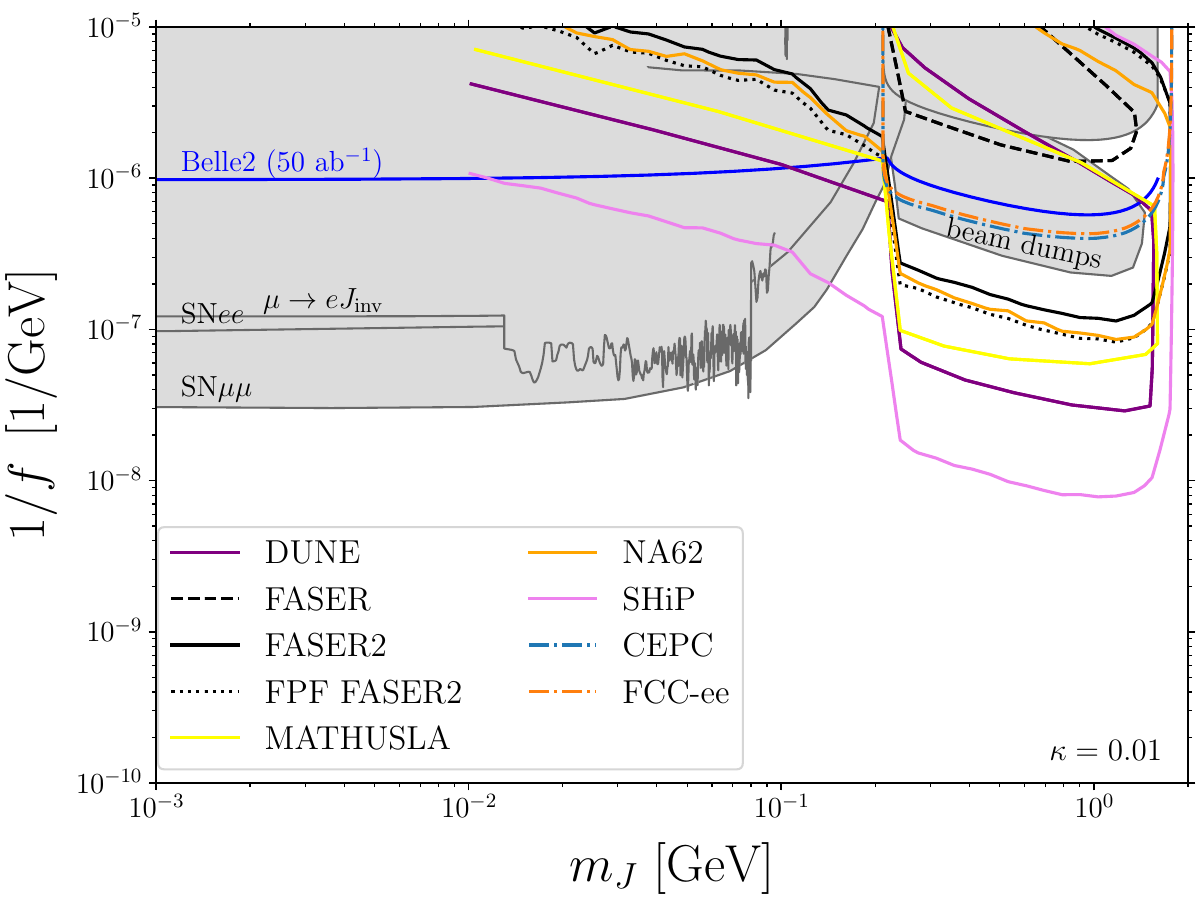}
          \hspace{0.22cm}
          \includegraphics[scale=0.43]{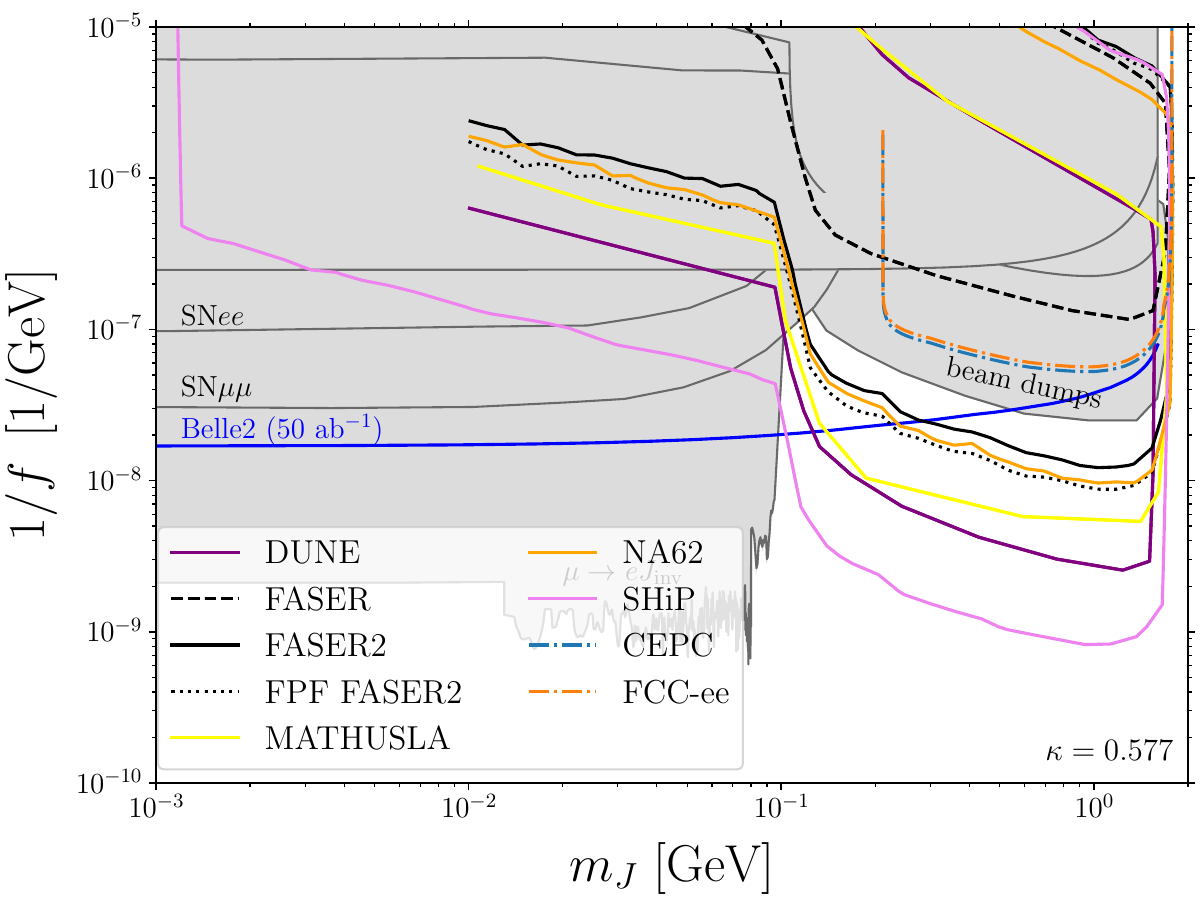}
          \caption{
                  Same as Fig.~\ref{fig:sensitivity_plot_1}, but for the complex texture defined in~\eqref{eq:K_benchmarks_2}, with $\kappa=0.01$ (left) and $\kappa=0.577$ (right). 
                  The diagonal entries are fixed at $K^d = 8\pi^2 v/f$, while $\kappa$ controls the magnitude of all off-diagonal (purely imaginary) LFV entries and hence the branching fractions $\mathrm{BR}(\tau^\pm\to\ell^\pm J)$ and, when kinematically allowed, the decay width $\Gamma(J\to e^\pm\mu^\mp)$.
                  }
\label{fig:sensitivity_plot_3}
\end{figure*}
\begin{figure*}[tb]
      \centering
          \includegraphics[scale=0.41]{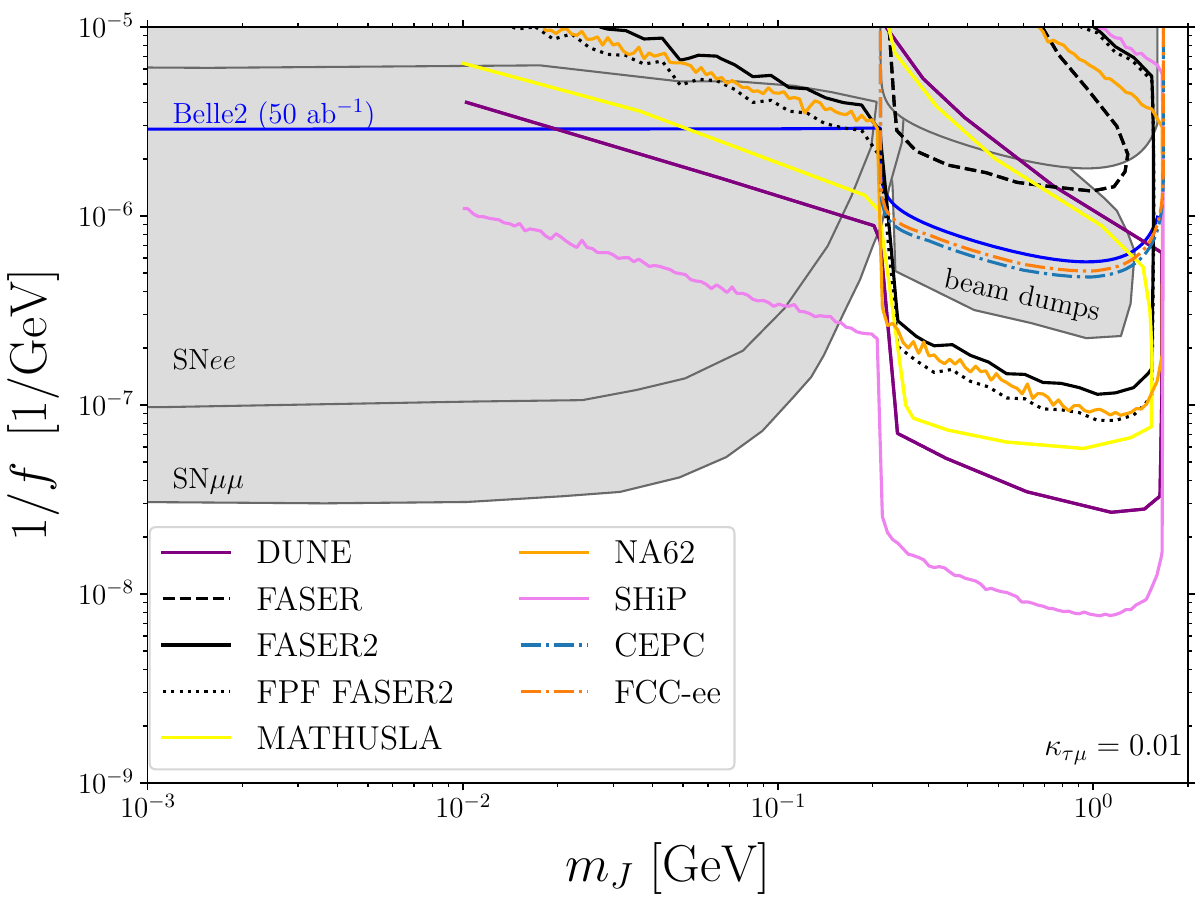}
          \hspace{0.22cm}
          \includegraphics[scale=0.41]{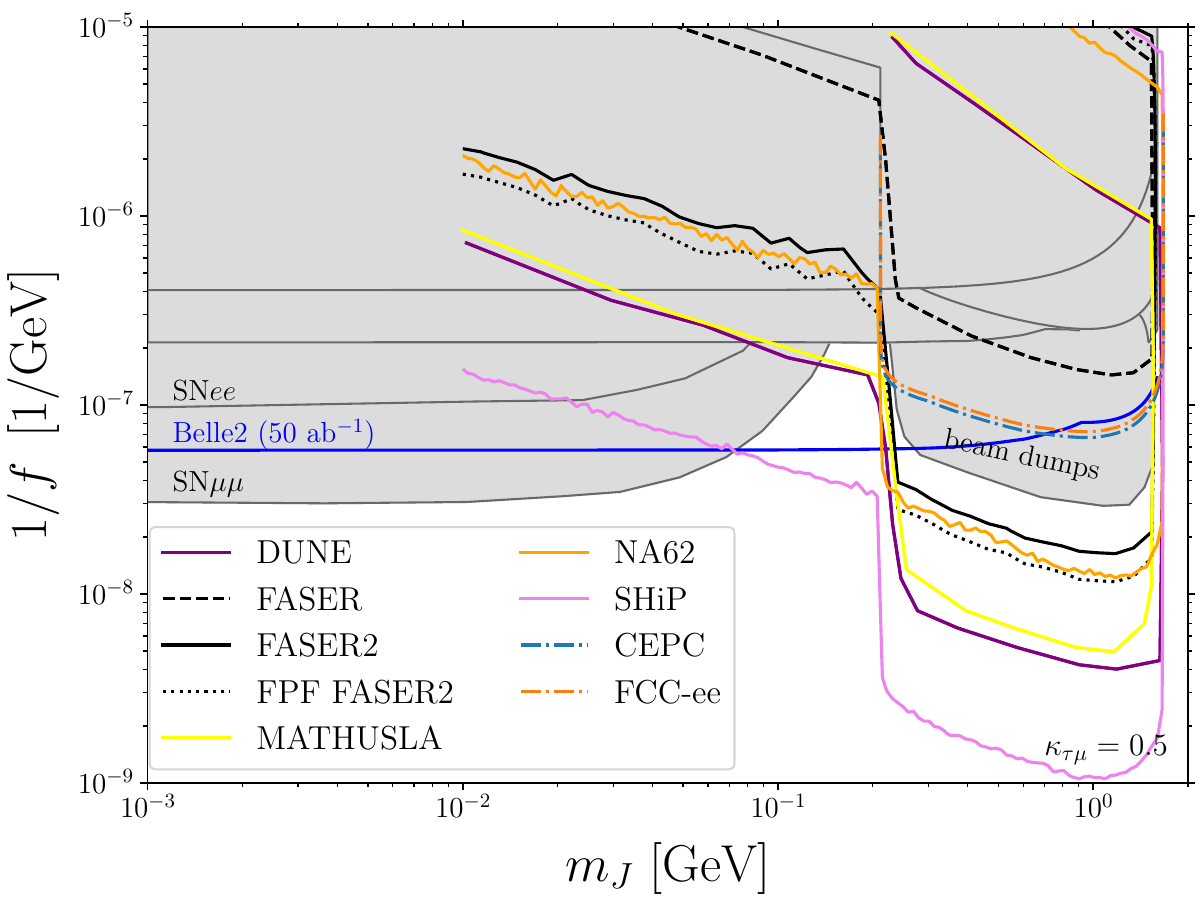}
          \caption{
                  Same as Fig.~\ref{fig:sensitivity_plot_1}, but for the single-flavor $K_{\mu\tau}$ texture defined in~\eqref{eq:K_benchmarks_1}, with $\kappa_{\mu\tau}=0.01$ (left) and $\kappa_{\mu\tau}=0.5$ (right). 
                  The diagonal entries are fixed at $K^d = 8\pi^2 v/f$, so that $\kappa_{\mu\tau}$ controls the magnitude of the $\tau$-$\mu$ LFV coupling and hence $\mathrm{BR}(\tau^\pm\to\mu^\pm J)$. The two panels illustrate the transition from a hierarchical regime, in which the off-diagonal entry is suppressed, to a near-anarchical regime with comparable diagonal and off-diagonal couplings.
                  }
\label{fig:sensitivity_plot_2}
\end{figure*}

\paragraph{Majoron decay widths and lifetime.} 
In Fig.~\ref{fig:decay_widths}, we present the majoron partial decay widths as a function of its mass for the three anarchical benchmarks, fixing $f=1\,$GeV (the general case scales as $(1\,\text{GeV}/f)^2$). 
The corresponding rest-frame decay length $c\tau$ is shown in Fig.~\ref{fig:ctau} for $f=10^7\,$GeV, a representative value probed by IF experiments at $m_J\sim 1\,$GeV---see the discussion below.

In the mass range probed here, $200\,\mathrm{MeV}\lesssim m_J \lesssim m_\tau - m_e \simeq 1.78\,$GeV, the dominant decay modes are the kinematically accessible lepton pairs---both FC ($e^+e^-$, $\mu^+\mu^-$) and FV ($\mu^\pm e^\mp$). 
For the benchmarks shown, the hadronic ($J\to 3\pi$, computed via~\eqref{eq:J_to_pi0_pi0_pi0}--\eqref{eq:J_to_pi0_piplus_piminus}) and diphoton decay modes are subdominant over the mass range relevant to tau production.
We checked this conclusion and the contributions from $\eta\pi\pi$, $\eta'\pi\pi$, and channels with kaons using the public code \textsc{ALPaca}~\cite{Alda:2025nsz} at representative benchmark points.
For $m_J \gtrsim 2\,$GeV, decays into nucleon pairs and inclusive hadronic final states (\eqref{eq:Gamma_J_had}) become dominant, but these masses exceed $m_\tau$.
For $K_{\rm anarch}$, the majoron partial decay widths are shown in Fig.~\ref{fig:decay_widths}, with a qualitatively similar pattern for the single-flavor (\eqref{eq:K_benchmarks_1}) and CP-violating (\eqref{eq:K_benchmarks_2}) benchmarks.

\paragraph{Majoron production mechanism and the role of LFV.} 
The LFV decays $\tau^\pm\to \mu^\pm J$ and $\tau^\pm\to e^\pm J$, induced at one loop by the off-diagonal entries of the matrix $K$, provide the dominant majoron production channel in all detectors considered.
The copious $\tau^\pm$ flux from $D_s/D \to \tau\nu_\tau$ decays in these facilities, combined with the unsuppressed LFV branching fractions (cf.~\eqref{eq:C_leptons}), yields efficient majoron production in the mass window $200\,\text{MeV} \lesssim m_J \lesssim m_\tau - m_e$. 
Subdominant production modes—--three-body meson decays, multi-body lepton decays, proton bremsstrahlung, and direct decays of on-shell $Z$ bosons---were found to result in negligible event rates compared to the $\tau^\pm$ LFV channel.
We have also considered majoron production from muon FV decays $\mu^\pm \to e^\pm J$ at FASER2, exploiting the high-energy ($E_\mu>100\,$GeV) muon beam passing through the detector~\cite{FASER:2021mtu,Ariga:2023fjg}, 
\be
    N^{\mathrm{FASER2}}_{\mu^\pm} = 4\times 10^{10}
\,.
\label{eq:N_mu}
\ee
However, since these probe the same low-mass, low-coupling regime as supernova cooling bounds~\cite{Calibbi:2020jvd,Li:2025beu}, the astrophysical constraints dominate in all scenarios considered. The results for this channel are shown only for the $R_J = 10$ benchmark shown in the right panel of Fig.~\ref{fig:sensitivity_plot_1}.

\paragraph{Systematic uncertainties.} 
The dominant source of systematic uncertainty in our beam dump and LHC far-detector projections is the modeling of forward charm meson production ($D_s^\pm,\, D^\pm$ spectra), which affects the $\tau^\pm$ flux and hence the majoron yield. 
We employed \textsc{Pythia\,8} (Monash tune) as our baseline generator and cross-checked with \textsc{SIBYLL} for LHC energies, finding qualitatively similar $\tau^\pm$ spectra. 
Residual uncertainties from the choice of fragmentation model, generator settings, and forward-production cross section can affect the signal yield by up to a factor of a few~\cite{Fieg:2023kld,Schubert:2024hpm}; an illustrative comparison of production models and their impact on the projected sensitivities is presented in Appendix~A of Ref.~\cite{Ema:2025bww}.
However, owing to the logarithmic scaling of the sensitivity contours with the number of signal events, these uncertainties translate into only modest shifts of the boundaries in the $(m_J,\,1/f)$ plane. The approximate scaling is $(1/f)_{max} \propto \log^{1/2}(N_{\tau^\pm})$ in the short-lived regime and $(1/f)_{min} \propto N_{\tau^\pm}^{-1/4}$ in the long-lived regime.

\paragraph{Sensitivity projections.} 
In Figs.~\ref{fig:sensitivity_plot_1}--\ref{fig:sensitivity_plot_2}, we present the projected sensitivity contours in the $(m_J,\,1/f)$ plane for all benchmarks listed in Sec.~\ref{sec:Coupling_structure}.
For beam dump experiments and LHC far detectors, we include both FC and FV charged-lepton final states, which extends the sensitivity below the dimuon threshold.
The CEPC and FCC-ee projections use $J\to\mu^+\mu^-$ and the reconstruction and residual-background assumptions specified below \eqref{eq:Zfactory_tau_yield}.
The beam dump and LHC far-detector projections correspond to $N_{\rm sig}=3$,  approximately a $95\%$ confidence-level sensitivity under negligible-background assumptions.
On the other hand, for NuCal, the electromagnetic and dimuon selections each yielded one event, with expected backgrounds of $0.3$ and $0.7$ events and detection efficiencies of $70\%$ and $80\%$, respectively, for visible energies above $10\,\mathrm{GeV}$~\cite{Blumlein:2013cua}. Therefore, we adopt a conservative threshold of $4.7$ selected signal events for each channel.

Among past beam dump experiments, we find that CHARM~\cite{CHARM:1985anb} provides the strongest existing limit, followed by BEBC~\cite{WA66:1985mfx} and NuCal~\cite{Blumlein:1990ay}; all three surpass the ArgoNeuT bound~\cite{Bertuzzo:2022fcm}. 
We denote the resulting combined exclusion region as ``beam dumps'' (gray shading). 
Among currently operating experiments, the NA62~\cite{NA62:2017rwk} projection reaches $f \sim 3\times 10^8\,$GeV for $m_J \simeq 1\,$GeV, while FASER~\cite{FASER:2018eoc}, thanks to the higher energy of the LHC beam, extends the sensitivity toward larger masses, up to $m_J \lesssim m_\tau - m_e$.

Among the beam dump and forward-detector proposals, SHiP will provide the largest reach in $f$ in our projections, with its best sensitivity reaching up to $1/f\sim3\times10^{-10}\,\mathrm{GeV}^{-1}$ (equivalently $f\sim3\times10^9\,\mathrm{GeV}$) at $m_J\simeq 1\,\mathrm{GeV}$ for the anarchical benchmarks.
This leading sensitivity is driven by SHiP's combination of large exposure ($6\times 10^{20}$ protons on target during its 15-year run), relatively short baseline ($L_{\rm min}=33.7\,$m), and long decay volume ($\Delta = 50\,$m). DUNE offers complementary but weaker coverage due to its lower beam energy (120~GeV vs.\ 400~GeV), reaching $f \sim \text{few}\times 10^8\;\text{GeV}$ at $m_J\simeq 1\,$GeV. 
At the LHC, MATHUSLA's large off-axis geometric acceptance compensates for the transverse production kinematics of the majorons from $\tau^\pm$ decays, yielding a reach comparable to that of DUNE and exceeding that of FASER2 and FPF~FASER2, which are optimized for the highly forward regime. 
For the anarchical benchmarks, the ordering of beam dump and LHC far-detector searches by maximum reach in $f$ is: SHiP, DUNE, MATHUSLA, FPF~FASER2, NA62, and FASER2.

The displaced-dimuon searches at CEPC and FCC-ee provide complementary probes of the same LFV $\tau^\pm$-decay production mechanism, while the $\tau^\pm$ production does not suffer from large uncertainties.
Moreover, the acceptance for displaced-dimuon decays close to the $\tau^\pm$ decay vertex reduces the loss of short-lived majorons before the fiducial region, whereas beam dump experiments require the majoron to traverse a substantial baseline.
In the representative $R_J=1.01$, CP-violating $\kappa=0.01$, and single-flavor $\kappa_{\mu\tau}=0.01$ benchmarks, SHiP nevertheless reaches smaller $1/f$ on the long-lived boundary.
For comparison, at $m_J=1\,\mathrm{GeV}$ the $Z$-factory projections reach $f\simeq(2$--$3)\times10^7\,\mathrm{GeV}$ for $R_J=1.01$, and $f\simeq(2$--$3)\times10^6\,\mathrm{GeV}$ for the two $\kappa=0.01$ textures.
The CEPC and FCC-ee curves nearly coincide because we use the same acceptance and reconstruction model. Their small separation reflects both the different $\tau^\pm$ yields and the different assumed background counts.
While the reconstruction efficiencies and residual backgrounds require a dedicated study at CEPC and FCC-ee, unlike the beam dump and forward-detector yields, these projections do not depend on the modeling of the forward charm production.
Finally, on the long-lived boundary, reducing a constant reconstruction efficiency from unity to $0.5$ or $0.1$ weakens the reach in $1/f$ by factors of approximately $1.19$ or $1.78$, respectively.

\paragraph{Benchmark dependence.} 
The qualitative sensitivity pattern is similar within each class of the $K$-matrix benchmarks, since in all cases $J$ is produced predominantly via a single process ($\tau^\pm$ decays). 
Nevertheless, the detailed reach of a given experiment can vary by up to an order of magnitude in $1/f$ across benchmarks, reflecting the strong dependence of the displaced-decay probability (\eqref{eq:prob_decay}) on the majoron lifetime---exponential suppression in the short-lived regime and $1/d$ scaling in the long-lived regime---as well as the quadratic dependence of the production rate on the relevant off-diagonal entries of the $K$ matrix.

The opposite trends visible in Fig.~\ref{fig:sensitivity_plot_1} compared to Figs.~\ref{fig:sensitivity_plot_3}--\ref{fig:sensitivity_plot_2} arise from the different parameterizations.
In the anarchical ansatz (\eqref{eq:K_anarch}), the off-diagonal couplings governing $\tau^\pm\to\ell^\pm J$ production are fixed to unity. 
Increasing $R_J$ leaves the production couplings unchanged and enhances the diagonal decay couplings, shortening the majoron lifetime and shifting the sensitivity contour to smaller $1/f$.
Conversely, for the single-flavor textures (\eqref{eq:K_benchmarks_1}) and CP-violating benchmarks (\eqref{eq:K_benchmarks_2}), the diagonal entries are fixed and the off-diagonal coupling $\kappa$ is varied. 
For the single-flavor textures, the majoron lifetime remains dominated by the diagonal couplings in the mass range considered. 
For the complex texture, $\kappa$ also controls $J\to e^\pm\mu^\mp$ and can significantly affect the lifetime. 
The sensitivity therefore depends on both the production rate, which scales as $\kappa^2$, and the majoron lifetime.
Finally, we note that the benchmark $R_J = 2/3$, employed in Ref.~\cite{Bertuzzo:2022fcm}, yields a $K$ matrix that is not positive semi-definite (cf.\ discussion below~\eqref{eq:K_anarch}) and therefore lies outside the physical parameter space of the majoron model. We retain it for comparison, but emphasize that only the $R_J = 1.01$ and $R_J = 10$ benchmarks satisfy positive definiteness.

\paragraph{Complementary constraints.} 
In Figs.~\ref{fig:sensitivity_plot_1}--\ref{fig:sensitivity_plot_2}, we overlay existing constraints and projections from several independent probes, both terrestrial LFV searches and astrophysical bounds.
Since such limits have been thoroughly discussed in a recent study of an axially coupled ALP, Ref.~\cite{Ema:2025bww}, we follow their discussion and use the digitized data in the GitHub repository accompanying~\cite{Ema:2025bww} to recast the limits from axially coupled ALP into majoron.

Supernova bounds: using the cooling bound~\cite{Li:2025beu,Carenza:2021pcm}, we recast the bounds separately from FC couplings (SN-$ee$ and SN-$\mu\mu$) and from FV couplings (SN-$e\mu$). We note that the trapping-regime (large-coupling) boundary of these exclusion regions may be weaker than indicated by standard estimates~\cite{Fiorillo:2023ytr,Fiorillo:2023cas}, and that supernova observables other than the total cooling luminosity can yield complementary or even stronger constraints~\cite{Fiorillo:2025sln,Telalovic:2024cot}.
A dedicated reassessment for the singlet majoron is beyond the scope of this work, dedicated to IF searches, but is not expected to alter our main conclusions, since the IF searches studied here target precisely the parameter region where astrophysical bounds lose sensitivity.
Indeed, while astrophysical observables provide the strongest existing limits for $m_J \lesssim 200\,$MeV, they lose sensitivity once $m_J$ exceeds the proto-neutron-star core temperature, $T_{\rm core}\sim 30$--$50\,$MeV.

LFV searches: we recast the existing limits from $\mu\to e + J_{\rm inv.}$~\cite{TWIST:2014ymv,PIENU:2020loi,Derenzo:1969za,Bilger:1998rp}, $\mu\to e +\gamma$~\cite{Bauer:2021mvw}, $\tau^\pm$ lifetime~\cite{ParticleDataGroup:2024cfk}, 
$\tau\to \mu +J_{\mathrm{inv.}}$~\cite{ARGUS:1995bjh,Belle:2025bpu, Belle-II:2022heu}, 
$\tau\to \mu +J_{\mathrm{vis.}}$~\cite{BaBar:2010axs,Hayasaka:2010np,ATLAS:2016jts,CMS:2020kwy,LHCb:2014kws,Belle-II:2024sce}; we suppress these labels in figures for clarity.
We also display projected sensitivities from Belle~II (see the blue solid lines in Figs.~\ref{fig:sensitivity_plot_1}--\ref{fig:sensitivity_plot_2}), assuming an integrated luminosity of $50$ ab$^{-1}$~\cite{RodriguezPerez:2019nhw}.

\paragraph{Gap-filling role of IF searches.} 
The sensitivity projections presented here demonstrate that beam dump experiments, forward-physics detectors, and $Z$ boson factories fill a critical gap in the intermediate mass window $200\,\text{MeV} \lesssim m_J \lesssim 1\,$GeV, between supernova cooling bounds, that lose sensitivity at $m_J \gtrsim T_{\rm core}$, and $B$-factory or dedicated LFV searches. 
Therefore, IF experiments provide complementary, often leading, probes of the benchmark coupling structures in this mass range.

\section{Conclusions}
\label{sec:conclusions}
In this work, we have evaluated existing limits and future sensitivities to a majoron with mass below $m_\tau$ at IF experiments.
We use representative effective coupling textures with the loop-induced charged-lepton interactions correlated through the positive-semidefinite matrix $K$. 
For each texture, we calculate LFV production rates, branching fractions, and lifetime of the majoron, and use these results as inputs to MC simulations of its signatures at all experiments listed in Table~\ref{tab:experiments}.

Our main findings can be summarized as follows.
Utilizing data from the past beam dump experiments BEBC, CHARM, and NuCal, and performing MC simulation of the $\tau^\pm$ spectra, majoron production, and decay-in-volume probabilities, we obtain substantially stronger constraints on majoron coupling strength than the existing ArgoNeuT bound~\cite{Bertuzzo:2022fcm}.

The structure and correlations of the majoron couplings, together with the positivity condition on the $K$ matrix are essential for consistent interpretation of the different textures.
In particular, the relation between axial and vector couplings, as well as the presence of couplings to gauge bosons, differ significantly from those of a purely axially coupled ALP~\cite{Ema:2025bww}.
While the anarchical, single-flavor LFV, and complex textures give a common framework for comparing majoron searches, their quantitative sensitivities can differ by up to an order of magnitude in the effective scale $f$.
This variation arises from the interplay between LFV production, the majoron lifetime, and the available decay channels.

The physical requirement of $K\succeq 0$ restricts the allowed coupling ratios.
For example, for the anarchical ansatz, the $R_J=2/3$ benchmark violates this condition, and we supplement it with the near-boundary benchmark $R_J=1.01$.

Among the beam-dump and LHC far-detector proposals considered, SHiP will provide  the largest projected reach in $f$, probing effective scales of order $10^9\,\mathrm{GeV}$ near $m_J=1\,\mathrm{GeV}$. 
DUNE, FASER2, and MATHUSLA provide complementary coverage through their different beam energies, baselines, and geometries. 
The calculated sensitivities extend into regions of the $(m_J,1/f)$ plane left unconstrained by the supernova and dedicated LFV bounds shown here, particularly for $m_J\simeq0.2\text{--}1\,\mathrm{GeV}$. The extent of this coverage depends strongly on the coupling texture and majoron decay length.

Looking ahead, several experimental developments will provide important complementary coverage.
Direct searches for $\tau^\pm\to\ell^\pm J$ at Belle~II~\cite{RodriguezPerez:2019nhw} will probe the branching fractions governing the majoron yield, searches for $\mu\to e\gamma$~\cite{MEGII:2018kmf} will constrain other LFV combinations, while the proposed Super Tau-Charm Facility~\cite{Ai:2025xop,Jiang:2025nie} could extend the reach further through high-statistics $\tau^\pm$ production.
On the theoretical side, improved modeling of forward charm production---the dominant systematic uncertainty of beam-dump projections---will sharpen the sensitivity forecasts. 
Finally, searches at $Z$-boson factories, CEPC and FCC-ee, avoid the uncertainty from forward charm production and could provide complementary probes of tau-produced majorons, with greater reach than the LHC forward detector FASER.
Together, these efforts will further probe the GeV-scale majoron.

\begin{acknowledgments}
The authors gratefully acknowledge the valuable discussions and insights provided by the members of the China Collaboration of Precision Testing and New Physics.
We thank Julian Heeck and Edoardo Vitagliano for useful comments. 
KJ thanks Taehyun Jung for discussions during the early stage of this work.

This work was supported by the National Natural Science Foundation of China (NNSFC) under grants No.~12335005, No.~12575118, and the Special funds for postdoctoral overseas recruitment, Ministry of Education of China. 
KJ was supported by the IBS under the project code, IBS-R018-D1.
\end{acknowledgments}

\bibliographystyle{utphys}

\bibliography{bibliography}

\end{document}